\documentclass[twocolumn,showpacs,prb,amsmath,amssymb,letterpaper]{revtex4}
\usepackage{graphicx}
\usepackage{bm}

\DeclareMathOperator{\Max}{Max}

\DeclareMathOperator{\diag}{diag}

\begin{document}

\title{Dangling-bond spin relaxation and magnetic 1/$f$ noise from the
  amorphous-semiconductor/oxide interface: Theory} \author{Rogerio de
  Sousa} 
\altaffiliation{Current address: Department of
  Physics and Astronomy, University of Victoria, Victoria, BC V8W 3P6, Canada.} 
\affiliation{Department of Physics, University of California,
  Berkeley, California 94720, USA} 

\date{\today}

\begin{abstract}
  We propose a model for magnetic noise based on spin-flips (not
  electron-trapping) of paramagnetic dangling-bonds at the
  amorphous-semiconductor/oxide interface.  A wide distribution of
  spin-flip times is derived from the single-phonon cross-relaxation
  mechanism for a dangling-bond interacting with the tunneling
  two-level systems of the amorphous interface.  The temperature and
  frequency dependence is sensitive to three energy scales: The
  dangling-bond spin Zeeman energy ($\delta$), as well as the minimum
  ($E_{\rm{min}}$) and maximum ($E_{\rm{max}}$) values for the energy
  splittings of the tunneling two-level systems. At the highest
  temperatures, $k_BT\gg \Max{(\delta, E_{\rm{max}})}$, the noise
  spectral density is independent of temperature and has a $1/f$
  frequency dependence.  At intermediate temperatures, $k_BT\ll
  \delta$ and $E_{\rm{min}}\ll k_BT\ll E_{\rm{max}}$, the noise is
  proportional to a power law in temperature and possesses a $1/f^p$
  spectral density, with $p=1.2-1.5$. At the lowest temperatures,
  $k_BT\ll \delta$, or $k_BT\ll E_{\rm{min}}$, the magnetic noise is
  exponentially suppressed.  We compare and fit our model parameters
  to a recent experiment probing spin coherence of antimony donors
  implanted in nuclear-spin-free silicon [T.  Schenkel {\it et al.},
  Appl.  Phys.  Lett.  {\bf 88}, 112101 (2006)], and conclude that a
  dangling-bond area density of the order of $10^{14}$~cm$^{-2}$ is
  consistent with the data.  This enables the prediction of single
  spin qubit coherence times as a function of the distance from the
  interface and the dangling-bond area density in a real device
  structure.  We apply our theory to calculations of magnetic flux
  noise affecting SQUID devices due to their Si/SiO$_2$ substrate.
  Our explicit estimates of flux noise in SQUIDs lead to a noise
  spectral density of the order of $10^{-12}\Phi_{0}^{2}
  (\rm{Hz})^{-1}$ at $f=1$~Hz. This value might explain the origin of
  flux noise in some SQUID devices.  Finally, we consider the
  suppression of these effects using surface passivation with
  hydrogen, and the residual nuclear-spin noise resulting from a
  perfect silicon-hydride surface.
\end{abstract}
\maketitle
%

\section{Introduction}

Our physical understanding of spin relaxation in semiconductors plays
a crucial role in the current development of spin-based
electronics\cite{zutic04} and spin-based quantum
computation.\cite{kane98} One question that received little or no
attention so far is related to magnetic noise in semiconductor devices
and nanostructures. Magnetic noise from impurities and other defects
at the interface may be the dominant source of spin phase relaxation
(decoherence) for implanted donor electrons\cite{schenkel06} or
nuclear spins\cite{ladd05} in isotopically purified silicon. Moreover,
because Si/SiO$_2$ and other amorphous oxide interfaces are used as
the substrate for sensitive SQUID
magnetometers,\cite{wellstood87,yoshihara06,clarke07} the spin
relaxation of magnetic impurities at the substrate might explain the
observed magnetic flux noise in these devices.

\begin{figure}
\includegraphics[width=3in]{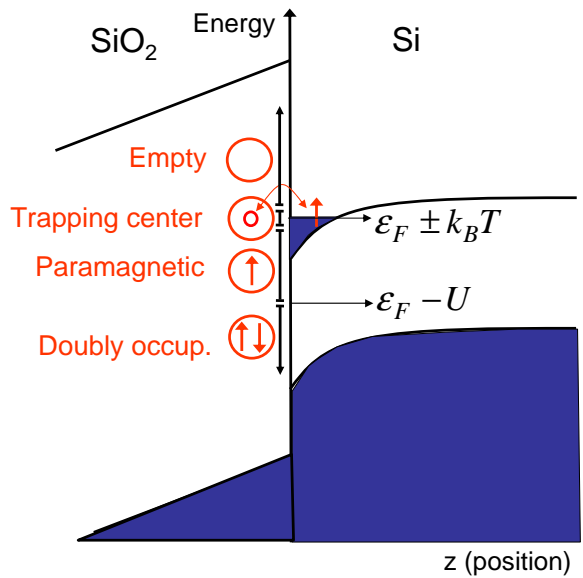}
\caption{(Color online) Band diagram for a Si/SiO$_2$ interface.
  Dangling bonds with energy much larger than $\epsilon_F$ are empty;
  DBs with energy in the interval $(\epsilon_F-k_BT,\epsilon_F+k_BT)$
  are trapping-centers for interface conduction electrons, responsible
  for charge, current, and magnetic noise.  DBs with energy in the
  interval $(\epsilon_F-U,\epsilon_F-k_BT)$ are singly occupied
  (paramagnetic), and hence contribute exclusively to magnetic noise.
  DBs with energy less than $\epsilon_F-U$ are doubly occupied and do
  not contribute to any kind of noise.}
\label{fig1}
\end{figure}

One universal characteristic of silicon devices is the presence of an
insulating interface, usually an oxide, separating the metallic gate
from the semiconductor. It is known for a long time that these
interfaces are rich in dangling-bond type defects (also denoted
``$P_b$ centers'') which can be detected using spin resonance
techniques.  These studies have established a wide distribution of
dangling-bond (DB) energy levels, spanning almost the whole
semiconductor energy gap, with each DB characterized by a large
on-site Coulomb energy $U\sim 0.5$~eV.\cite{lenahan,gerardi86} When
the dangling-bond (DB) energy level falls within $k_BT$ of the
interface Fermi level, it acts as a trapping-center and leads to the
well known $1/f$ charge and current noise for interface conduction
electrons.\cite{kogan96} Nevertheless at low temperatures the area
density for trapping-center DBs is only a tiny fraction of the area
density for \emph{paramagnetic} DBs.  For example, at $T=5$~K this
fraction is only $k_BT/U\sim 10^{-3}$ (Fig.~\ref{fig1}).  As a
consequence, the magnetic noise due to paramagnetic DBs is at least a
factor of $U/k_BT \gg 1$ larger than magnetic noise generated by
electron trapping, provided the paramagnetic DBs have a non-zero
spin-flip rate (Magnetic noise due to electron trapping is discussed
in appendix~\ref{trappingcenters}).

The spin relaxation rate for dangling-bond type defects depends
crucially on the non-crystalline nature of amorphous
compounds.\cite{kurtz80,lyo80,askew86} However, a detailed theoretical
study of the magnetic field and temperature dependence of this effect
has not been done.  In this article we present a general theory of
dangling-bond spin-lattice relaxation in amorphous materials, and show
that the noise created by the magnetic dipolar field of an ensemble of
dangling-bonds has the $1/f$ frequency dependence at high
temperatures.  We fit our theory to a recent experiment probing spin
coherence of antimony donors implanted in nuclear-spin-free
silicon\cite{schenkel06} in order to estimate our model parameters.

\begin{figure}
\includegraphics[width=3in]{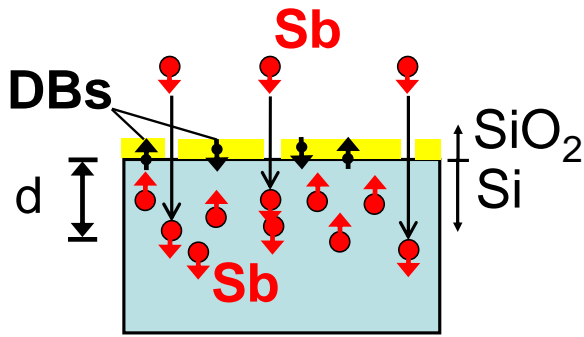}
\caption{(Color online) How to detect low frequency magnetic noise
  using electron spin resonance. A low density of antimony (Sb) donor
  impurities is implanted in a Si/SiO$_2$ sample using an ion gun, and
  the distribution of Sb donors is determined using secondary ion mass
  spectroscopy.  Next, a Hahn echo decay experiment is performed on
  the Sb spins.\cite{schenkel06} The Hahn echo decay envelope is
  directly related to magnetic noise produced by e.g. dangling bonds
  at the interface, see Eq.~(\ref{splus}).}
\label{fig2}
\end{figure}

We exploit the important relationship between phase coherence of a
localized ``probe'' spin (e.g. the implanted Sb spins in
Ref.~\onlinecite{schenkel06}) and its environmental magnetic noise
(Fig.~\ref{fig2}).  The coherence decay envelope of a ``probe'' spin
measured by a class of pulse spin resonance sequences is directly
related to a frequency integral over magnetic noise times a filter
function.\cite{desousa06} This allows us to interpret pulse spin
resonance experiments of localized spins as sensitive detectors of
magnetic noise in nanostructures.  The spin qubit phase coherence is a
local probe of low frequency magnetic noise. 
The same ideas apply equally well to experiments probing the coherent
dynamics of superconducting devices.\cite{nakamura02,martinis03}

An important step towards this characterization was given recently, by
the report of the the first measurements of spin echo decay in silicon
implanted with an ultra-low dose of antimony donors ($\sim
10^{11}$~cm$^{-2}$).\cite{schenkel06} Two samples were reported,
120~KeV and 400~KeV, with low and high implant energy respectively.
The former leads to a donor distribution closer to the interface, see
Table I.

\begin{table}
\begin{center}
\begin{tabular}{c c c c c}
\hline\hline
Sample & Interface & Peak depth [nm]  & $T_1$ [ms]& $T_2$ [ms]\\\hline
120~KeV & Si/SiO$_{2}$ & 50  & $15\pm 2$ & $0.30\pm 0.03$\\\hline
120~KeV & Si-H & 50  & $16\pm 2$ & $0.75\pm 0.04$\\\hline
400~KeV & Si/SiO$_{2}$ & 150  &  $16\pm 1$ & $1.5\pm 0.1$ \\\hline
400~KeV & Si-H & 150 &$14\pm 1$ & $2.1\pm 0.1$ \\\hline
\hline
\end{tabular}
\caption{Spin relaxation data\cite{schenkel06} taken at
  5.2~K for antimony donor electron spins implanted in isotopically
  purified silicon. $T_1$ was measured using inversion recovery ESR,
  while $T_2$ is the 1/e decay of Hahn echo. For each sample, data
  was taken for the untreated oxidized surface (SiO$_{2}$) and for the
  passivated surface, treated with hydrofluoric acid in order to
  obtain a hydrogen terminated surface. The data clearly indicates
  that (1) donors close to the surface have lower spin coherence
  times $T_2$ but the same spin-flip time $T_1$; (2) Surface
  passivation leads to a sizable increment in $T_2$, but no
  change in $T_1$.}
\label{tableI}
\end{center}
\end{table}

Table I provides experimental evidence that the surface leads to
additional mechanisms for donor spin phase fluctuation and magnetic
noise.  These mechanisms seem to contribute exclusively to the phase
coherence time ($T_2$) but not to the spin-flip time ($T_1$) of the Sb
donors, therefore the associated noise spectrum should be low
frequency in nature (with a high frequency cut-off much smaller than
the spin resonance frequency).

Here we consider the mechanisms of magnetic noise that might be
playing a role in these experiments. For a Si/SiO$_2$ interface we
show that dangling-bond spin-flips play a dominant role.  A
dangling-bond (DB) is a paramagnetic defect usually associated with an
oxygen vacancy in the Si/SiO$_2$ interface. These point defects are
generically denoted ``$P_b$ centers'' with chemical structure
represented by $\textrm{Si}_3\equiv
\textrm{Si}\cdot$.\cite{lenahan,gerardi86} There is yet no
experimental or theoretical studies of spin relaxation times ($T^{{\rm
    DB}}_{1}$) for DBs at the Si/SiO$_2$ interface.  Nevertheless a
systematic study of DB spin relaxation in bulk amorphous silicon was
carried out in the 1980's.\cite{kurtz80,askew86} The measured DB spin
relaxation rate was found to increase as a power law on temperature,
$1/T^{\rm{DB}}_{1}\propto T^{n}$ with an anomalous exponent $n=2-4$
dependent on the sample preparation method. At $T=5$~K and $B=0.3$~T
the typical $T^{\rm{DB}}_1$ was in the range $0.1-1$~ms.\cite{askew86}

At first it seems puzzling that the dangling-bond spin would relax in
such a short time scale at the lowest temperatures.  The typical $T_1$
of localized electron spins in crystalline silicon (e.g. phosphorus
donor impurities) is almost a thousand seconds in the same
regime.\cite{feher59} This happens due to the weak spin-orbit coupling
in bulk crystalline silicon.  However, dangling-bonds in
non-crystalline silicon are coupled to unstable structural defects,
and this fact seems to explain their short
$T_1$.\cite{kurtz80,askew86} These structural defects behave as
tunneling two level systems strongly coupled to lattice vibrations
(phonons).  Each time a tunneling two level system (TTLS) undergoes a
phonon-induced transition, the DB spin feels a sudden shift in its
local spin-orbit interaction, which may be quite large because the
TTLS is associated with a local reordering of the atomic positions of
the non-crystalline material.  As a consequence, the DB spin may flip
each time the TTLS switches.  Remarkably, this cross-relaxation
process remains effective even at zero magnetic field because it does
not involve a Kramers conjugate pair (in contrast to spin-flips
without a simultaneous TTLS switch).

We develop this theory further in order to incorporate the
exponentially wide TTLS parameter distribution typical of amorphous
materials.  As a result, we find that the magnetization of an
initially polarized ensemble of DB spins will undergo non-exponential
relaxation in time. Our theory of dangling-bond spin-lattice
relaxation and magnetic noise is based on an effective Hamiltonian
approach, allowing us to draw generic conclusions about the frequency,
temperature, and magnetic field dependence of spin-noise in a variety
of amorphous materials.  For example, our results apply equally well
to the magnetic noise produced by $E'$ centers in bulk SiO$_2$,
another well studied dangling-bond.  Other materials of relevance to
our work are the bulk Al$_2$O$_3$ (sapphire), and Al/Al$_2$O$_3$ and
Si/Si$_3$N$_4$ interfaces, whose paramagnetic dangling-bonds/magnetic
impurities are yet to be characterized experimentally.

Our results are of particular importance to magnetic flux noise in
SQUID devices, whose microscopic origin is a longstanding puzzle (for
a review see section IV-G of Ref.~\onlinecite{weissman88}).  In
section~\ref{squid} we apply our results to calculations of flux noise
due to DBs within the area enclosed by the SQUID loop, and show that
this contribution might explain some of the available flux noise
measurements.

It is possible to considerably reduce the dangling-bond area density
using a surface passivation technique. For example, the application of
hydrofluoric acid to the Si/SiO$_2$ surface removes dangling-bonds by
covering the surface with a monolayer of hydrogen atoms.  Recently,
Kane and collaborators fabricated a field-effect-transistor using a
passivated Si(111)H surface, and demonstrated record high electron
mobility.\cite{eng05} Nevertheless, the large density of hydrogen
nuclear spins might be an important source of magnetic noise. The
nuclear spins are constantly fluctuating due to their mutual dipolar
coupling.  In section~\ref{nuclearsd} we consider calculations of
magnetic noise due to a hydrogen terminated Si(100)H surface. We use
the same theory previously developed for Hahn echo decay of a
phosphorus impurity in bulk doped natural
silicon.\cite{witzel05,desousa06} We show that the Hahn echo decay in
a Si(100)H surface has many peculiarities, including a special crystal
orientation dependence for the donor $T_2$ times that may be used as
the fingerprint for detecting this source of noise experimentally.

\section{Relationship between magnetic noise and phase relaxation 
  in pulse spin resonance experiments: Electron spin as a local probe
  of magnetic noise}

Consider the following model Hamiltonian for the interaction of a
localized spin
with a noisy environment,
\begin{equation}
{\cal H} = \frac{1}{2}\gamma_{e} B \sigma_z + 
\hat{\bm{\eta}}(t)\cdot \bm{\sigma}.
\label{hmodel}
\end{equation}
Here $\bm{\sigma}=(\sigma_{x},\sigma_{y},\sigma_{z})$ is the vector of
Pauli matrices denoting the state of the electron spin being probed by
a pulse spin resonance experiment (henceforth called the donor spin -
e.g. the Sb spins in Ref.~\onlinecite{schenkel06}), $\gamma_{e} B$ is
the spin Zeeman frequency in an applied external magnetic field $B$,
and $\gamma_{e}=g e/(2 m_e c)$ is a gyromagnetic ratio for the
electron spin [for a group V donor impurity such as P or Sb,
$\gamma_{e}\approx 1.76\times 10^{7}$~(sG)$^{-1}$ is close to the free
electron value]. Note that Eq.~(\ref{hmodel}) was divided by $\hbar$
so that energy is measured in units of frequency.  Each component of
the vector $\hat{\bm{\eta}}=(\hat{\eta}_x,\hat{\eta}_y,\hat{\eta}_z)$
is an operator modeling the magnetic environment (the DB or other
impurity spins) surrounding the donor spin.  The simplest way to
describe the time evolution of the spin's magnetization $\langle
\bm{\sigma}\rangle$ is the Bloch-Wangsness-Redfield approach, which
assumes $\langle \bm{\sigma}\rangle$ satisfies a first order
differential equation in time.  The decay rate for $\langle
\sigma_z\rangle$ is then given by
\begin{equation}
\frac{1}{T_1}=\frac{\pi}{2}\sum_{q=x,y}\left[
\tilde{S}_{q}(+\gamma_{e} B)+\tilde{S}_{q}(-\gamma_{e} B)\right],
\end{equation}
with the environmental noise spectrum defined by
\begin{equation}
\tilde{S}_{q}(\omega)=\frac{1}{2\pi}\int_{-\infty}^{\infty}
\textrm{e}^{i\omega
t} \langle \hat{\eta}_{q}(t)\hat{\eta}_{q}(0)\rangle dt.
\label{noisedef}
\end{equation}
Note that the energy relaxation time $T_1$ for the donor spin is determined
by the noise at $\omega=\pm \gamma_{e} B$, that is just a statement of
energy conservation. Within the Bloch-Wangsness-Redfield theory 
the spin's transverse magnetization
($\langle\sigma_{+}\rangle=\langle \sigma_x+i\sigma_y\rangle/2$) 
decays exponentially with the rate 
\begin{equation}
\frac{1}{T^{*}_{2}}=\frac{1}{2T_1}+ \pi \tilde{S}_{z}(0),
\label{t2}
\end{equation}
where we added a $*$ to emphasize this rate refers to a free induction
decay (FID) experiment. The Bloch-Wangsness-Redfield approach leads to a simple
exponential time dependence for all spin observables. Actually this is
not true in many cases of interest, including the case of a group V
donor in bulk silicon where this approximation fails completely (for
Si:P the observed Hahn echo decay fits well to
$\textrm{e}^{-\tau^{2.3}}$ in many
regimes).\cite{witzel05,tyryshkin06} The problem lies in the fact that
the Bloch-Wangsness-Redfield theory is based on an infinite time limit
approximation, that averages out finite frequency fluctuations. Note
that $T^{*}_{2}$ differs from $T_1$ only via static noise,
$\tilde{S}_z(0)$ in Eq.~(\ref{t2}). A large number of spin resonance
sequences, most notably the Hahn echo are able to remove static noise
completely. 

We may develop a theory for spin decoherence that takes into account
low frequency fluctuations in the semiclassical regime $\hbar\omega\ll
k_BT$, when $\tilde{S}_z(-\omega)=
\textrm{e}^{-\hbar\omega/k_BT}\tilde{S}_z(\omega)\approx
\tilde{S}_z(\omega)$. The spin coherence envelope may be calculated in
the pure dephasing limit ($\hat{\eta}_x=\hat{\eta}_y=0$), with the
assumption that $\hat{\eta}_z\rightarrow \eta_z$ is distributed
according to Gaussian statistics.  For derivations and discussions on
the applicability of this theory, we refer to
Ref.~\onlinecite{desousa06}.  A similar method in the context of
superconducting qubits was proposed in Ref.~\onlinecite{martinis03}.
The final result is a direct relationship between phase coherence and
magnetic noise according to
\begin{equation}
\left|\langle \sigma_{+}(t)\rangle\right|
=\exp{\left[-\int_{-\infty}^{\infty}d\omega\;\tilde{S}_{z}(\omega)
{\cal F}(t,\omega)\right]},
\label{splus}
\end{equation}
with ${\cal F}(t,\omega)$ a
filter function that depends on the
particular pulse spin resonance sequence. 
For a free induction decay experiment
($\pi/2-t-\rm{measure}$) 
we have
\begin{equation}
{\cal F}_{\rm{FID}}(t,\omega)=
\frac{1}{2}\frac{\sin^2{\left(\omega t/2\right)}}{\left(\omega/2\right)^2}, 
\label{ffid}
\end{equation}
while for the Hahn echo ($\pi/2-\tau-\pi-\tau-\rm{measure}$) the
filter function becomes
\begin{equation}
{\cal F}_{\rm{Hahn}}(2\tau,\omega)=
\frac{1}{2}\frac{\sin^4{\left(\omega \tau/2\right)}}{\left(\omega/4\right)^2}.
\label{fecho}
\end{equation}
Note that in the limit $t\rightarrow \infty$ Eq.~(\ref{ffid}) becomes
$\pi\delta(\omega)t$, recovering the Bloch-Wangsness-Redfield result
Eq.~(\ref{t2}). The Hahn echo filter function satisfies ${\cal
  F}_{\rm{Hahn}}(2\tau,0)=0$, showing that it filters out terms
proportional to $\tilde{S}_z(0)$ in spin evolution. This is equivalent
to the well known removal of inhomogeneous broadening by the spin
echo. Any pulse spin resonance sequence containing instantaneous
$\pi/2$ or $\pi$-pulses can be described by Eq.~(\ref{splus}). Another
important example is the class of Carr-Purcell sequences used for
coherence control ($\pi/2-[\tau-\pi-\tau-\rm{echo}]_{\rm{repeat}}$).

\section{Dangling-bond spin relaxation: Direct vs. cross-relaxation}

\begin{figure}
\includegraphics[width=3in]{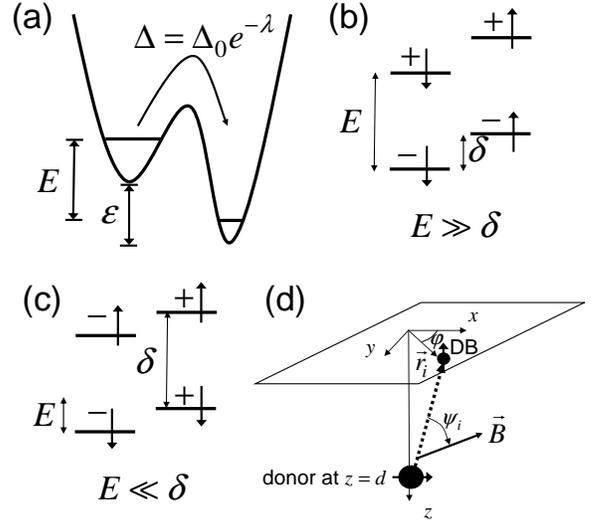}
\caption{(a) Effective double well potential for the tunneling two
  level system (TTLS). (b,c) Energy level structure for a
  dangling-bond spin (DB) coupled to a TTLS, for (b) $E\gg \delta$ and
  (c) $E\ll \delta$. (d) Coordinate system for the interaction of
  a dangling-bond located at $\bm{r}_i$ with the donor spin. 
$\psi_i$ denotes the angle formed by the donor-DB vector (dashed) and
  the external $B$ field.}
\label{fig3}
\end{figure}

The presence of an inversion center in crystalline Si leads to weak
spin-orbit coupling and extremely long spin relaxation times.
The $T_1$ for localized donor electrons in crystalline silicon can
reach thousands of seconds at low temperatures.\cite{feher59} This is
in contrast to spin-lattice relaxation of dangling-bonds in various
forms of amorphous silicon where instead $T^{{\rm DB}}_{1}$ was found
to range between one and a hundred milliseconds at the lowest
temperatures ($T=0.3-4$~K).\cite{askew86} The proposed theoretical
explanation was that DB spin relaxation happens due to its coupling to
phonon-induced transitions of tunneling two level systems (TTLS) in
the amorphous material.\cite{kurtz80} The TTLSs are thought to be
structural rearrangements between groups of atoms, that can be modeled
by a double well potential [see Fig.~\ref{fig3}(a)]. The TTLS assumption is
able to explain several special properties of amorphous materials at
low temperatures.\cite{galperin89} The DB spin couples to the TTLSs
either through spin-orbit or hyperfine interaction, both of which are
modulated by the TTLS transition. Note that the presence of a TTLS
breaks the crystal inversion symmetry.

We start by developing the theory of phonon-induced transitions for the
TTLS,\cite{jackle72} and the associated cross-relaxation of the DB
spin. The Hamiltonian for a TTLS reads
\begin{equation}
{\cal H'_{\rm{TTLS}}}=\frac{1}{2}\left(
\begin{array}{cc}
\epsilon & \Delta\\
\Delta & -\epsilon
\end{array}
\right).\label{h0p}
\end{equation}
The energy scale $\epsilon$ is a double well asymmetry, while $\Delta=
\Delta_{0}\textrm{e}^{-\lambda}$ is the tunneling matrix element
between the states [$\lambda$ is related to the barrier height and its
thickness, see Fig.~\ref{fig3}(a)]. After diagonalizing Eq.~(\ref{h0p}) we
obtain ${\cal H}_{\rm{TTLS}}=\diag{\{E/2,-E/2\}}$, with
$E=\sqrt{\epsilon^2+\Delta^2}$ (for notational clarity we prime the
Hamiltonians in the non-diagonal basis).  The coupling
to phonons can be obtained by expanding the parameter $\epsilon$ to
first order in the phonon strain operator, 
\begin{equation}
\hat{u}=i\sum_{\bm{q}} \sqrt{\frac{\hbar}
{2\rho V \omega_{q}}}|\bm{q}|\left(
a_{q} \textrm{e}^{i\bm{q}\cdot \bm{r}}
+a^{\dag}_{q}\textrm{e}^{-i\bm{q}\cdot\bm{r}}\right),
\end{equation}
leading to $\epsilon\rightarrow \epsilon +\epsilon'\hat{u}$. Below we
average over TTLS parameters with $\epsilon\gg \Delta$, so to be
consistent we must assume the deformation parameter $\Delta'=0$.
Applying this expansion to Eq.~(\ref{h0p}) and transforming to the
diagonal basis we get
\begin{equation}
{\cal H}_{\rm{TTLS-ph}} = \frac{\epsilon'\hat{u}}{2E}
\left(
\begin{array}{cc}
\epsilon & -\Delta\\
-\Delta & -\epsilon
\end{array}
\right).\label{httlsph}
\end{equation}
Using Fermi's golden rule for dissipation into a phonon bath
${\cal H}_{\rm{ph}}=\sum_{q} \hbar\omega_{q}a^{\dag}_{q}a_{q}$,
we find that the transitions from
$+E/2$ to $-E/2$ and vice-versa are given by
\begin{eqnarray}
r_{+}&=&aE\Delta^{2} \left[n_{\rm{ph}}(E)+1\right],\label{rp}\\
r_{-}&=&aE\Delta^{2} n_{\rm{ph}}(E),\label{rm}
\end{eqnarray}
with phonon occupation number
\begin{equation}
n_{\rm{ph}}(E)=\frac{1}{\textrm{e}^{E/k_BT}-1}.
\end{equation}
In Eqs.~(\ref{rp}),~(\ref{rm}) the parameter $a$ depends on the
material density $\rho$,sound velocity $s$, and deformation potential
$\epsilon'$ [$a=(8\pi |\epsilon'|^2\hbar^4 \rho s^5)^{-1}$].  The DB
spin Zeeman energy is denoted by ${\cal H}_{\rm{DB}}=\hbar\gamma_e B
S^{\rm{DB}}_{z}$. To simplify the notation we define $\delta\equiv
\hbar\gamma_e B$ as the DB spin Zeeman energy.  The coupling of the DB
spin to the TTLS may be derived directly from the spin-orbit
interaction ${\cal H}_{\rm{so}}=\alpha \bm{S}^{\rm{DB}}\cdot
(\bm{E}\times \bm{p})$, where $\bm{S}^{\rm{DB}}$ is the DB spin operator,
$\bm{p}$ is the DB orbital momentum, and $\bm{E}$ a
local electric field. After averaging over the coordinate states, the
resulting effective Hamiltonian becomes directly proportional to the
magnetic field, a consequence of time reversal symmetry.\cite{vanvleck40}
For simplicity, we assume that $\bm{E}$ is perpendicular to the
interface,\cite{note3} and that the spin-orbit energy fluctuates by a certain
amount $A\times\delta$ when the TTLS switches. This leads to the
following effective Hamiltonian in the non-diagonal basis
\begin{equation}
{\cal H'}_{\rm{TTLS-DB}}= \frac{A\delta}{2}\left(S^{\rm{DB}}_{+}
+S^{\rm{DB}}_{-}\right)
\left(
\begin{array}{cc}
1 & 0\\
0 & -1
\end{array}
\right),\label{ttlsspindiag}
\end{equation}
where $S^{\rm{DB}}_{\pm}$ are raising and lowering operators for the DB
spin. The dimensionless constant $A$ will play the role of a small parameter
in our theory.  Transforming to the diagonal basis we get
\begin{equation}
{\cal H}_{\rm{TTLS-DB}}= \frac{A\delta}{2E}\left(S^{\rm{DB}}_{+}
+S^{\rm{DB}}_{-}\right)
\left(
\begin{array}{cc}
+\epsilon & \Delta\\
\Delta & -\epsilon
\end{array}
\right).\label{ttlsspin}
\end{equation}
As a result of Eq.~(\ref{ttlsspin}), the DB-TTLS eigenstates
are admixtures between spin up and down.  We may still label the
eigenstates by their spin quantum number, provided we think of
$\uparrow$ ($\downarrow$) as having a large projection onto the pure
spin up (down) state. 
The four level
structure is shown in Fig.~\ref{fig3}(b) in the limit $E\gg \delta$ and in
Fig.~\ref{fig3}(c) for $E\ll \delta$.

The total Hamiltonian is given by 
\begin{equation}
{\cal H}={\cal H}_{\rm{TTLS}}
+{\cal H}_{\rm{DB}}
+{\cal H}_{\rm{TTLS-DB}}
+{\cal H}_{\rm{ph}}
+{\cal H}_{\rm{TTLS-ph}}.
\end{equation}
Note that the first three contributions denote
the discrete TTLS-DB states (a four-level system), the fourth is the
energy bath (a continuum of phonon states) and the fifth is the
coupling between the TTLS-DB to the phonon bath. The eigenstates of
the first three contributions may be calculated using perturbation
theory, and the transition rates are straightforward to compute. The
``direct'' relaxation rate, corresponding to a DB spin-flip 
\emph{with the TTLS state unchanged} 
is given by 
\begin{equation}
D_{\pm\uparrow\rightarrow \pm \downarrow}=\frac{a}{4}\frac{\Delta^4
  A^2}{E^2 \left(E^2-\delta^2\right)^{2}}\delta^{5}
\left[n_{\rm{ph}}(\delta)+1\right],
\label{directrs}
\end{equation}
with $[n_{\rm{ph}}(\delta)+1]\rightarrow n_{\rm{ph}}(\delta)$ for the
reverse rate $D_{\pm\downarrow\rightarrow \pm \uparrow}$.  Note that
Eq.~(\ref{directrs}) is proportional to $\Delta^4$, reflecting the
fact that a direct spin-flip may only occur together with a virtual
transition to an excited orbital state.\cite{vanvleck40,feher59} In
our case this virtual transition is a ``double-switch'' of the TTLS,
hence $D\propto\Delta^4$ [terms independent of $\Delta$ in
Eq.~(\ref{directrs}) cancel exactly. This general feature of a direct
spin-flip process is referred to as ``van Vleck
cancellation'',\cite{vanvleck40} giving a simple explanation of why
direct spin-flip rates are generally weak]. Moreover,
Eq.~(\ref{directrs}) vanishes at $B=0$ in accordance with time
reversal symmetry (the direct process couples a Kramers pair).

The ``cross''-relaxation rates, whereby the \emph{DB spin flips
  simultaneously with a TTLS switch} are given by
\begin{eqnarray}
\Gamma_{-\downarrow}&=&a \left|M_{+}\right|^{2}
\left(E+\delta\right)
n_{\rm{ph}}(E+\delta),\label{wmdpu}\\
\Gamma_{+\uparrow}&=&
a\left|M_{+}\right|^{2}
\left(E+\delta\right)
\left[n_{\rm{ph}}(E+\delta)+1\right],\label{wpumd}\\
\Gamma_{-\uparrow}&=&
a\left|M_{-}\right|^{2}
\left(E-\delta\right)
n_{\rm{ph}}(E-\delta),\label{wmupd}\\
\Gamma_{+\downarrow}&=&
a\left|M_{-}\right|^{2}
\left(E-\delta\right)
\left[n_{\rm{ph}}(E-\delta)+1\right],\label{wpdmu}
\end{eqnarray}
where the sub-indexes label the level that the system is exiting, for example 
$\Gamma_{+\uparrow}\equiv \Gamma_{+\uparrow\rightarrow
  -\downarrow}$. Note that the final state is obtained from the
initial state by changing the sign of the TTLS and flipping the DB spin.
The matrix element $M_{\pm}$ is defined by
\begin{equation}
M_{\pm}=\frac{A\epsilon\Delta}{E^{2}}\left[
\left|E\pm \delta\right|+\delta\right].
\label{mpm}
\end{equation}

Remarkably, this cross-relaxation process is \emph{not a transition
  between Kramers conjugate states}.  As a result, the rates are
qualitatively different from the direct process, particularly due to
their magnetic field ($\delta$) and TTLS energy ($E$) dependence.  At
low temperatures ($k_BT\ll \delta$), the direct rate always scales as
$D\propto \delta^5$.\cite{vanvleck40,feher59} In contrast, the
cross-relaxation rate has two distinct behaviors, depending whether
$E\gg \delta$, or $E\ll \delta$. For $E\gg \delta$, $M_{\pm}\approx
A\epsilon\Delta/E$, and the $\Gamma$'s are independent of magnetic
field. For $E\ll \delta$ we get instead $M_{\pm}\approx 2\delta
A\epsilon\Delta/E^2$, and $\Gamma\propto \delta^3$ in contrast to the
$\delta^5$ scaling of the direct rate.

Of extreme importance to our theory is to note that whenever the
energy scales $E$ and $\delta$ are well separated, the direct rates
are much smaller than the cross relaxation rates. For $E\gg \delta$ we
have $D/\Gamma\sim (\Delta/\epsilon)^2 (\delta/E)^5$, while for $E\ll
\delta$ we have $D/\Gamma\sim (E/\epsilon)^{2}(\Delta/\delta)^{2}$.
The typical assumption for amorphous semiconductors is $\Delta\ll
\epsilon,\delta$ and $E\approx \epsilon$.\cite{kogan96}
In this regime the direct rates are substantially weaker than
the cross relaxation rates, except at the resonance point $E=\delta$.
It is useful to list simple expressions for the cross-relaxation rates
in the two most physically relevant regimes considered in this work.
For low magnetic field $\delta\ll k_BT$, $E\gg \delta$ but with 
$E/k_BT$ arbitrary we have simply
\begin{equation}
\Gamma_{\pm\uparrow}=\Gamma_{\pm\downarrow}\equiv \Gamma_{\pm}\approx A^{2}r_{\pm}.
\label{gar}
\end{equation}
Hence when the spin-orbit coupling parameter satisfies $A\ll 1$, the
cross-relaxation spin-flips are much less frequent than the 
spin-preserving TTLS switching events. The opposite high magnetic field regime
with $E\ll \delta$ and $E\ll k_BT$ with $\delta/k_BT$ arbitrary leads to
\begin{equation}
\Gamma_{+\uparrow}\approx \Gamma_{-\uparrow}\equiv \Gamma_{\uparrow}\approx 
4a\frac{\delta^3A^2\Delta^2}{E^2}[n_{\rm{ph}}(\delta)+1],
\label{gad}
\end{equation}
with the reverse rate $\Gamma_{\downarrow}$ given by
$[n_{\rm{ph}}(\delta)+1]\rightarrow n_{\rm{ph}}(\delta)$. Note that
these $\Gamma_{\uparrow\downarrow}$ rates are still much larger than
the direct rates, since $D/\Gamma_{\uparrow\downarrow}\sim
(\Delta/\delta)^2$.

Finally, we discuss how the cross-relaxation rates are affected by the
presence of phonon broadening in a non-crystalline material.  In this
case we generalize our theory by including a complex part to the
phonon spectra, $\omega_q =sq +i\gamma_{\rm{ph}}$. The modified 
Eq.~(\ref{mpm}) becomes
\begin{eqnarray}
M_{\pm}&=&\frac{A\epsilon\Delta}{E^2}\left[
\frac{\delta^2+\gamma_{\rm{ph}}^{2}/2}{\delta^2+\gamma_{\rm{ph}}^{2}}
\sqrt{\left(E\pm\delta\right)^2+\gamma_{\rm{ph}}^{2}}\right.\nonumber\\ 
&&\left.+\frac{\delta}{\left|E\pm \delta\right|}\frac{\left(E\pm
\delta\right)^{2}
+\gamma_{\rm{ph}}^{2}/2}{\sqrt{\left(E\pm \delta\right)^{2}}+\gamma_{\rm{ph}}^{2}}
\right].
\end{eqnarray}
For amorphous Si we estimate $\gamma_{\rm{ph}}\sim 0.01
sq$.\cite{fabian96} For $E\gg \delta$, $\gamma_{\rm{ph}}\sim 0.01 E$
may be comparable to $\delta$, and we see that $M_{\pm}$ is reduced by
a factor of two, and an additional $B$ field dependence results.

\section{Dangling-bond spin relaxation: Ensemble average\label{defects}}

In order to evaluate the ensemble averages over TTLS parameters we
must first determine the time-dependent correlation function for the
four-level relaxation network described in Fig.~\ref{fig3}(b,c). Using the
notation of Eq.~(\ref{noisedef}), the magnetic dipolar field
produced by a single DB spin maps into a c-number
$\hat{\eta}_z=2h_{\rm{dip}}S_{z}^{\rm{DB}}\rightarrow h_{\rm{dip}}s_i
$, with $s_i=+1$ (DB spin up) or $s_i=-1$ (DB spin down). 
In the four-level system notation
$(+\downarrow,-\downarrow,+\uparrow,-\uparrow)$ the vector $\bm{x}$ of
dipolar fields assumes the values $\bm{x}=h_{\rm{dip}}(-1,-1,+1,+1)$.
In Appendix~\ref{calctimecor} we prove the convenient identity
\begin{equation}
S_z(t)=\left\langle \left[\eta_z(t)-\bar{\eta}_z\right]
\left[\eta_z(0)-\bar{\eta}_z\right]\right\rangle
=\bm{x}\cdot \bm{p}(t)\cdot \bm{x_{w}}.\label{stexp}
\end{equation}
Here $\bm{x_{w}}=(x_1 w_1,x_2 w_2,\ldots)$, with $w_i$ the equilibrium
probabilities for the i-th level of the DB+TTLS network. The matrix
$\bm{p}(t)=\textrm{e}^{-\bm{\Lambda}t}$ describes the occupation
probability for each level, and decays according to the relaxation
tensor $\Lambda$. Below we discuss the important analytic solutions
for $S_z(t)$ in the limit of small spin-orbit coupling, $A\ll 1$.

\subsection{Case $E\gg \delta$, $\delta\ll k_BT$, $E/k_BT$ arbitrary}

In this regime the TTLS and cross-relaxation rates are simply related
by Eq.~(\ref{gar}).  The time correlation function for the DB spin may
be calculated exactly from Eq.~(\ref{stexp}), but for simplicity we
show the result to lowest order in powers of $A$:
\begin{equation}
S_z(t) \approx h_{\rm{dip}}^{2}\left[
\Psi\textrm{e}^{-(r_++r_--\bar{\Gamma})t}+\left(1-\Psi\right)\textrm{e}^{-\bar{\Gamma}t}\right],
\label{ege}
\end{equation}
with a visibility loss given by 
\begin{equation}
\Psi=\frac{\tanh^2{(E/2k_BT)}}{\cosh^2{(E/2k_BT)}}A^4,
\label{visib}
\end{equation}
and a thermalized DB spin relaxation rate given by 
\begin{eqnarray}
\bar{\Gamma}&=&\frac{2r_{-}}{r_{+}+r_{-}}\Gamma_{+}+
\frac{2r_{+}}{r_{+}+r_{-}}\Gamma_{-}\nonumber\\
&\approx & 2 a A^2 \Delta^2
\frac{E}{\sinh{\left(E/k_BT\right)}},
\label{gammabar}
\end{eqnarray}
where we used $\epsilon\approx E$. Interestingly, Eq.~(\ref{ege})
shows that DB spin relaxation happens in two stages: In the first
stage the DB spin decays abruptly to a small visibility loss $\Psi$,
with a rate set by the TTLS switch. During this first stage the TTLS
levels $\pm E/2$ achieve thermal equilibrium.  In the second stage the
DB spin relaxes fully with a much slower ``thermalized''
cross-relaxation rate $\bar{\Gamma}$. For $A\ll 1$ we may drop the
$\Psi\propto A^4$ contribution to Eq.~(\ref{ege}).

The theory developed above can be generalized to a single DB coupled
to an ensemble of TTLSs, provided the TTLSs are not coupled to each
other. In this case the rate Eqs.~(\ref{gammabar})~and~(\ref{dbr2})
are generalized to a sum of rates $\Gamma_i$ relating to the i-th
TTLS. Each exponential in Eq.~(\ref{ege}) becomes
$\sim \textrm{e}^{-\sum_i\Gamma_i t}$. This happens whenever the
DB+TTLS network can be separated into disconnected four level
subspaces as in Fig.~\ref{fig3}(b,c).


We now proceed to average over disorder realizations of the amorphous
material. We assume the following two-parameter distribution
\begin{equation}
P(\lambda,E)=\frac{\bar{P}v}{\lambda_{\rm{max}}}\left(
\frac{E}{E_{\rm{max}}}\right)^{\alpha},
\label{Eldist}
\end{equation}
for $\lambda\in [0,\lambda_{\rm{max}}]$, and $E\in
[E_{\rm{min}},E_{\rm{max}}]$; $P(\lambda,E)=0$ otherwise.  Note that
the uniform distribution in $\lambda$ leads to a broad distribution of
TTLS tunneling parameters $\Delta=\Delta_0 \textrm{e}^{-\lambda}$.  To
our knowledge there are no estimates available for
$\bar{P},E_{\rm{max}},E_{\rm{min}}$ close to an interface, only for
bulk SiO$_2$. For the latter material the energy density of TTLSs per
unit volume $\bar{P}$ has been estimated as $\bar{P}=
10^{20}-10^{21}\rm{eV}^{-1}$~cm$^{-3}$, and typical values for the
TTLS energy range are $E_{\rm{min}}/k_B\sim 0.1$~K, and
$E_{\rm{max}}/k_B\sim 10$~K.\cite{galperin89} Here we introduce a new
parameter $v$ with units of volume, denoting the effective range for
TTLSs to couple to a each DB spin (for a SiO$_2$ layer of 10~nm we
estimate $v\sim 10^3$~nm$^3$).
The exponent $\alpha$ is material dependent: While $\alpha\approx 0$
seems to be appropriate for bulk SiO$_2$,\cite{galperin89} it was
found that bulk amorphous Si can be described by $\alpha = 0.1-0.4$ or
$\alpha= 1.2-1.5$ depending on sample preparation method (See
Ref.~\onlinecite{askew86} and section~\ref{compaskew} below). 

The average number of TTLSs coupled to each DB spin is given by
\begin{equation}
{\cal N}=\int d\lambda\int dE\; P(\lambda,E)\approx
\frac{\bar{P}vE_{\rm{max}}}{\alpha+1}.
\label{nttls}
\end{equation}
This is also the number of thermally activated TTLSs at high
temperatures, $k_BT\gg E_{\rm{max}}$. For lower temperatures
satisfying $E_{\rm{min}}\ll k_BT\ll E_{\rm{max}}$, Eq.~(\ref{nttls})
is divided by $\cosh^2{(E/2k_BT)}$, leading to
\begin{equation}
{\cal N}_{T} \approx \bar{P}vE_{\rm{max}}\left(\frac{2k_BT}{E_{\rm{max}}}\right)^{1+\alpha}.
\label{nttlsT}
\end{equation}
This is the number of \emph{thermally activated TTLSs interacting with
  each DB spin}. For extremely low temperatures $k_BT\ll E_{\rm{min}}$
this number will be exponentially small.

We now turn to computations of the ensemble averaged DB spin
relaxation rate, $\langle \bar{\Gamma}\rangle$.  At shorter times
satisfying $\bar{\Gamma}_{\rm{Max}}t\ll 1$, the DB spin magnetization
$\left\langle S_{z}(t)\right\rangle$ decays linearly in
time.\cite{note0} The rate for this linear decay is equivalent to the
$\left\langle 1/ T_{1}^{\rm{DB}}\right\rangle$ rate measured for bulk
amorphous silicon samples in Ref.~\onlinecite{askew86}. This is given by 
\begin{equation}
\left\langle \bar{\Gamma}\right\rangle =
\left\langle \frac{1}{T_{1}^{{\rm DB}}}\right\rangle =
\int d\lambda\int dE P(\lambda,\epsilon)\bar{\Gamma}(\lambda,E).
\label{avggammabar}
\end{equation}

At high temperatures $k_BT\gg E_{\rm{max}}$ Eqs.~(\ref{avggammabar})~and~(\ref{gammabar}) lead to
\begin{equation}
\left\langle \bar{\Gamma}\right\rangle = a A^2 \Delta_{0}^{2} k_BT \frac{{\cal N}}{\lambda_{\rm{max}}}.
\end{equation}
The average DB spin relaxation scales linearly with temperature times
the number of TTLSs surrounding the DB.

At lower temperatures satisfying $E_{\rm{min}}\ll k_BT\ll
E_{\rm{max}}$ we have instead
\begin{eqnarray}
\left\langle \bar{\Gamma}\right\rangle &=&
\frac{a A^2 \Delta_{0}^{2}\bar{P}v}{\lambda_{\rm{max}}E_{\rm{max}}^{\alpha}}\left(k_B T\right)^{2+\alpha}
\int_{0}^{\infty} dx \frac{x^{\alpha}}{\sinh^2{x}}\nonumber\\
&=& \frac{3a A^2 \Delta_{0}^{2}k_B T}{\lambda_{\rm{max}}2^{1+\alpha}} {\cal N}_T,
\label{t1dblow}
\end{eqnarray}
showing that the DB spin relaxation rate will scale proportional to
$T^{2+\alpha}$.

At the very lowest temperatures $k_BT\ll E_{\rm{min}}$ there are no
thermally activated TTLSs, therefore the mechanism of DB
cross-relaxation is exponentially suppressed.  Here other sources of
DB spin relaxation may dominate [e.g. direct relaxation as in
Eq.~(\ref{directrs})], or the DB spin may not relax at all within the
characteristic time scale of the experiment.

Askew {\it et al.} measured average DB relaxation rates in bulk
amorphous silicon at low temperatures ($T=0.3-5$~K).\cite{askew86}
Two different preparation methods, silicon implanted with $^{28}$Si,
and silicon sputtered in a substrate, led to the experimental fit
$\langle \bar{\Gamma} \rangle \propto T^{2.35}$. Two other
preparation methods, silicon implanted with $^{20}$Ne, and silicon
evaporated on a substrate led to $T^{3.3}$ and $T^{3.5}$ fits
respectively.  Two different values of the magnetic field were studied
($0.3$ and $0.5$~T), and no magnetic field dependence could be
detected.  The $T$ and $B$ dependence predicted by our model agrees
with experiment provided $\alpha=0.35$ for the $^{28}$Si implanted and
sputtered samples, and $\alpha=1.3,1.5$ for the $^{20}$Ne implanted
and the evaporated samples. It's perhaps expected that $\alpha$ is
different for each of these because the density of TTLSs should depend
on the way they were created. At high temperatures, the linear in $T$
behavior has been observed in amorphous silicon grown by
evaporation.\cite{gourdon81}


\subsection{Case $E\ll \delta$, $E\ll k_BT$, $\delta/k_BT$ arbitrary}

From Eq.~(\ref{gad}) and Eq.~(\ref{stexp}) 
we get
\begin{equation}
S_z(t)\approx \frac{h_{\rm{dip}}^{2}}{\cosh^2{(\delta/2k_BT)}}
\textrm{e}^{-\left(\Gamma_{\uparrow}+\Gamma_{\downarrow}\right)t}.\label{esd}
\end{equation}
For $E\ll \delta$ the DB relaxation rate becomes
\begin{equation}
\Gamma_{\uparrow}+\Gamma_{\downarrow}\approx 4aA^2
\frac{\Delta^2}{E^2}\delta^{3} \coth{\left(\frac{\delta}{2k_BT}\right)}.
\label{dbr2}
\end{equation}
Its ensemble average is given by
\begin{eqnarray}
\left\langle \Gamma_{\uparrow}+\Gamma_{\downarrow}\right\rangle &\approx&
\frac{2 a A^2 }{\lambda_{\rm{max}}}\left(\frac{\Delta_{0}^{2}}{E_{\rm{max}}E_{\rm{min}}}\right)
\left(\frac{E_{\rm{max}}}{E_{\rm{min}}}\right)^{\alpha}\nonumber\\
&&\times\left(\frac{1+\alpha}{1-\alpha}\right)
{\cal N}\delta^3 \coth{\left(\frac{\delta}{2k_BT}\right)},
\label{t1rateegd}
\end{eqnarray}
where we assumed $\alpha<1$. For $\alpha\geq 1$, the prefactor in
Eq.~(\ref{t1rateegd}) is modified, but the scaling $\propto {\cal
  N}\delta^{3}\coth{(\delta/2k_BT)}$ remains.

\subsection{Comparison to Ref.~\onlinecite{askew86}\label{compaskew}}

We now compare our results to the theoretical model proposed by Askew
{\it et al.}.\cite{askew86} In their Eq.~(5) the authors wrote the
expression for $\bar{\Gamma}$ in the $E\gg \delta$ regime using free
parameters $D$, $M$, $C$, $N$. In our work these are explicitly
related to microscopic parameters: $D=\epsilon'\epsilon/E$,
$M=-\epsilon'\Delta/E$, $C=\epsilon/(2E)$, $N=\Delta/(2E)$. In
Ref.~\onlinecite{askew86} it is claimed that when the inequality
$ND/E\gg -CM/\delta$ is satisfied, the average DB relaxation rate
scales as $\langle T^{{\rm DB}}_{1}\rangle^{-1}\propto
T^{2+\alpha}\delta^{0}$ (the so called Lyo and Orbach regime after
Ref.~\onlinecite{lyo80}). When this inequality is reversed, they
obtained $\langle T_{1}^{{\rm DB}}\rangle^{-1}\propto
T^{4+\alpha}\delta^{-2}$ (Kurtz and Stapleton regime,
Ref.~\onlinecite{kurtz80}). Nevertheless, our result shows that these
parameters are related by $ND=-CM>0$, so this inequality is equivalent
to $\delta\gg E$. Because Eqs.~(7)~and~(8) of
Ref.~\onlinecite{askew86} are based on two conflicting approximations, 
$\delta\gg E$ for the matrix element squared and $\delta\ll E$ for
the phonon density of states, their result needs to be
corrected.  We showed above that the average DB relaxation 
scales instead as $\delta^{3}\coth{(\delta/2k_BT)}$ for $\delta\gg E$ and
$T^{2+\alpha}\delta^{0}$ for $\delta\ll E$ (the latter holds for
$E_{\rm{min}}\ll k_BT\ll E_{\rm{max}}$. For high temperatures $k_BT\gg
E_{{\rm max}}$ we get $T\delta^{0}$). The corrected results are in
excellent qualitative agreement with the experimental data in
Ref.~\onlinecite{askew86}.

Ref.~\onlinecite{askew86} assumes $\epsilon=\Delta=E/\sqrt{2}$ and
averages $E$ according to a density $\sim E^{\alpha}$. This is in
contrast to our averaging prescription that assumes instead
$\Delta=\Delta_0\textrm{e}^{-\lambda}$, with
$\Delta_{0}<\epsilon_{\rm{min}}$ and as a consequence $\epsilon\approx
E$. We assume $\lambda$ is uniformly distributed and the $\epsilon$
density varies as $\sim \epsilon^{\alpha}$.  This assumption is
motivated by the wide distributions of TTLS relaxation rates observed
in glasses, and is usually employed to explain charge and current
noise in semiconductors.\cite{kogan96} As we show below, the broader
distribution of DB relaxation times leads to $1/f$ magnetic noise and
non-exponential relaxation for an ensemble of DBs.

\section{Magnetic noise \label{magnoise}}

The total noise power for each DB spin is independent of the specific
relaxation process and may be calculated exactly using elementary
Boltzman statistics. The noise must satisfy the following sum rule:
\begin{equation}
\int_{-\infty}^{\infty}\tilde{S}_z(\omega)d\omega =\langle \eta_{z}^{2}\rangle -\langle \eta_z\rangle^{2}
=\frac{\langle h_{\rm{dip}}^{2}\rangle}{\cosh^{2}{(\delta/2k_BT)}}. 
\label{sumrule}
\end{equation}
This shows that the noise spectrum is exponentially small in the high
magnetic field regime $\delta\gg k_BT$. For the opposite regime
$\delta\ll k_BT$ the total noise power
is independent of temperature. However, as we show below, the spectral
density $\tilde{S}_{z}(\omega)$ may be temperature dependent when its
upper frequency cut-off is temperature dependent.

\subsection{Case $E\gg \delta$, $\delta\ll k_BT$, $E/k_BT$ arbitrary}

In order to determine the noise spectrum, we must first extract the
distribution of relaxation rates $P(\bar{\Gamma})$ from
Eqs.~(\ref{gammabar})~and~(\ref{Eldist}). Under the assumption that
each DB spin is coupled to only one TTLS on average [i.e., ${\cal
  N}\sim 1$ or ${\cal N}_T\sim 1$, see
Eqs.~(\ref{nttls}),~(\ref{nttlsT})] we have
\begin{equation}
P(\Gamma')=\frac{\int d\lambda \int dE P(\lambda,E)\delta\left(\bar{\Gamma}(\lambda,E)-\Gamma'\right)}
{\int d\lambda \int dE P(\lambda,E)}.
\label{pgammadef}
\end{equation}
Note that this is normalized to one according to $\int d\Gamma'
P(\Gamma')=1$. It is straightforward to extend Eq.~(\ref{pgammadef})
to a larger number of TTLSs $E_1,E_2,\ldots$, but the explicit
calculation of $P(\Gamma')$ becomes difficult. Below we will derive
explicit results for the case of a DB spin coupled to a single TTLS on
average.

Using Eqs.~(\ref{gammabar}),~(\ref{Eldist}),~and~(\ref{pgammadef}) we may evaluate 
the integral over $\lambda$ explicitly:
\begin{eqnarray}
P(\Gamma')&=& \frac{1}{{\cal N}} \int dE 
P(0,E)\int d\lambda 
\frac{\delta[\lambda - \lambda_0 (E)]}{|\frac{d\bar{\Gamma}}{d\lambda}|_{\lambda=\lambda_0(E)}}\nonumber\\ 
&=& \frac{1}{2\Gamma'}\frac{1}{{\cal N}}
\int dE P(0,E)\nonumber\\
&&\times \theta \left[\frac{2aA^2 E}{\sinh{(E/k_BT)}}-\Gamma'\right].
\label{pgammaexpl}
\end{eqnarray}
Here $\lambda_0(E)$ is the solution of $\bar{\Gamma}(\lambda_0,E)=\Gamma'$.
The step function results from the fact that the delta function will
``click'' only when $\lambda_0(E)\in [0,\lambda_{\rm{max}}]$, or
simply $\Gamma'\leq 2aA^2E/\sinh{(E/k_BT)}$.

\subsubsection{High temperature,  $k_BT\gg E_{\rm{max}}$}

In this case the theta function in Eq.~(\ref{pgammaexpl}) is always
one for $\Gamma'\in
[\bar{\Gamma}_{\rm{min}},\bar{\Gamma_{\rm{max}}}]$, with
$\bar{\Gamma}_{\rm{max}}=2aA^2k_BT$ and
$\bar{\Gamma}_{\rm{min}}=\textrm{e}^{-2\lambda_{\rm{max}}}\bar{\Gamma}_{\rm{max}}$.
Therefore we have simply
\begin{equation}
P(\Gamma')=\frac{1}{2\lambda_{\rm{max}}\Gamma'},
\end{equation}
for $\Gamma'\in [\bar{\Gamma}_{\rm{min}},\bar{\Gamma_{\rm{max}}}]$,
and $P(\Gamma')=0$ otherwise. As a check, note that $\int
d\Gamma'P(\Gamma')=1$ implies the relationship
$\lambda_{\rm{max}}=\frac{1}{2}
\ln{\left(\frac{\bar{\Gamma}_{\rm{max}}}{\bar{\Gamma}_{\rm{min}}}\right)}$,
as expected.

The magnetic noise is given by
\begin{eqnarray}
\tilde{S}(\omega)&=& \langle h_{\rm{dip}}^{2}\rangle \int d\Gamma P(\Gamma)
\frac{\Gamma/\pi}{\omega^{2}+\Gamma^{2}}\nonumber\\
&=& \frac{\langle h_{\rm{dip}}^{2}\rangle}{4\lambda_{\rm{max}}}\frac{1}{|\omega|},
\label{somegaegd}
\end{eqnarray}
for $\bar{\Gamma}_{\rm{min}}<\omega<\bar{\Gamma}_{\rm{max}}$, and
$\tilde{S}(\omega)=0$ for $\omega>\bar{\Gamma}_{\rm{max}}$. For
$\omega<\bar{\Gamma}_{\rm{min}}$ it saturates at
$\tilde{S}(\bar{\Gamma}_{\rm{min}})$. \emph{Hence at the highest
temperatures we have temperature independent magnetic 1/$f$
  noise}.

The 1/$f$ frequency dependence shows that the average magnetization of
an ensemble of DB spins out of equilibrium will decay
non-exponentially with time $t$. At intermediate times satisfying
$\bar{\Gamma}_{\rm{max}}^{-1} \ll t\ll \bar{\Gamma}_{\rm{min}}^{-1}$,
we may show that the time correlation function (or equivalently the
ensemble average of the DB $z$-magnetization) satisfies\cite{kogan96}
\begin{equation}
\frac{\left\langle S^{\rm{DB}}_{z}(t)\right\rangle}{\left\langle
S^{\rm{DB}}_{z}(0)\right\rangle}\approx 1-\frac{C_{\rm{E}}+
\ln{(\bar{\Gamma}_{\rm{max}}t)}}{2\lambda_{\rm{max}}}.
\label{nonexponential}
\end{equation}
This expression is valid after neglecting terms ${\cal
  O}(1/\bar{\Gamma}_{\rm{max}}t)$. Here $C_{\rm{E}}=0.5772$ is the
Euler-Mascheroni constant.

\subsubsection{Intermediate temperature,  $E_{\rm{min}} \ll k_BT \ll E_{\rm{max}}$ \label{intermediatetemp}}

In this case Eq.~(\ref{pgammaexpl}) becomes
\begin{eqnarray}
P(\Gamma')&=&\frac{1}{2\Gamma'} \frac{1+\alpha}{\lambda_{\rm{max}}}\left(
\frac{k_BT}{E_{\rm{max}}}
\right)^{1+\alpha}\nonumber\\
&&\times\int_{0}^{x_{\rm{max}}} dx\; x^{\alpha} \theta\left(
\bar{\Gamma}_{\rm{max}}\frac{x}{\sinh{x}}-\Gamma'
\right).
\end{eqnarray}
The upper limit of the integral is determined from
$\frac{x}{\sinh{x}}=\Gamma'/\bar{\Gamma}_{\rm{max}}$. We solved this
equation numerically, and showed that the result is well approximated
by the analytic expression $x_{\rm{max}}\approx
\frac{3}{2}\left|\ln{\left(\frac{\Gamma'}{2\bar{\Gamma}_{\rm{max}}}\right)}\right|$.
Using this approximation we get
\begin{equation}
P(\Gamma')=\frac{1}{2\lambda_{\rm{max}}\Gamma'}\left(\frac{k_BT}{E_{\rm{max}}}\right)^{1+\alpha}
\left|
\frac{3}{2}\ln{\left(\frac{\Gamma'}{\bar{\Gamma}_{\rm{max}}}\right)}
\right|^{1+\alpha}.
\label{pgammalow}
\end{equation}
The distribution of relaxation rates has the same temperature
dependence as the number of thermally activated TTLSs [see
Eq.~(\ref{nttlsT})], and possesses an interesting logarithmic
correction with respect to the usual $1/\Gamma'$ behavior.  

\begin{figure}
\includegraphics[width=3in]{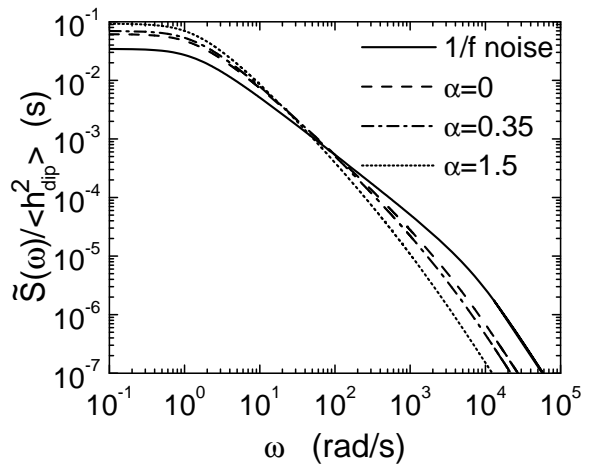}
\caption{Magnetic noise at intermediate temperatures $k_BT\gg \delta$
  and $E_{\rm{min}}\ll k_BT\ll E_{\rm{max}}$, for $\alpha=0,0.35,1.5$
  (TTLS energy density exponents). The distribution of relaxation
  rates [Eq.~(\ref{pgammalow})] contains a logarithmic correction,
  leading to $\tilde{S}(\omega)\propto 1/f^{p}$, with
  $p=1.2-1.5$.\label{lowTnoise}}
\end{figure}

The logarithm correction in Eq.~(\ref{pgammalow}) increases the weight
for smaller rates $\Gamma'$, at the expense of decreasing the weight
for higher rates. As a result the noise spectrum is better described
by a 1/$f^p$ relation, with $p>1$.  Fig.~\ref{lowTnoise} shows
numerical calculations of $\tilde{S}(\omega)$ for $\alpha=0,0.35,1.5$
(we assumed $\bar{\Gamma}_{\rm{min}}=1$~s$^{-1}$, and
$\bar{\Gamma}_{\rm{max}}=10^{4}$~s$^{-1}$). For $\alpha=0$, the noise
is described by a $1/f^{1.2}$ fit, while for $\alpha=1.5$ a fit of
$1/f^{1.5}$ is more appropriate. Therefore at intermediate
temperatures we have
\begin{equation}
  \tilde{S}(\omega)=
  \frac{\langle h_{\rm{dip}}^{2}\rangle}{4\lambda'_{\rm{max}}}
\left(\frac{k_BT}{E_{\rm{max}}}\right)^{1+\alpha}\frac{1}{|\omega|^{p}}.
\label{somegap}
\end{equation}
Note that $\lambda'_{\rm{max}}$ is determined from the normalization
condition $\int d\omega \tilde{S}(\omega)=\langle
h_{\rm{dip}}^{2}\rangle$ for given
$\bar{\Gamma}_{\rm{max}}/\bar{\Gamma}_{\rm{min}}$.


\subsubsection{Extremely low temperature,  $k_BT \ll E_{\rm{min}}$}

In this case $\bar{\Gamma}(\lambda,E)$ is exponentially suppressed,
and there will be no magnetic noise due to the DB+TTLS mechanism.  If
spin relaxation is dominated by the direct process
[Eq.~(\ref{directrs})], the noise spectra may still have the $1/f$
dependence. Otherwise paramagnetic DBs may not contribute to magnetic
noise at all.

\subsubsection{Calculation of $\langle h_{\rm{dip}}^{2}\rangle$}

Finally, we calculate the total noise power by averaging the DB
distribution over the interface plane.
We choose a coordinate system with origin at the interface immediately
above the donor spin. Define $d$ as the donor depth, and $r_i$,
$\phi_i$ the coordinates of the $i$th DB with respect to the
interface [see Fig.~\ref{fig3}(d)]. 
The dipolar frequency shift produced by a DB spin aligned along the same
direction as the donor spin is given by
\begin{equation}
(h_{\rm{dip}})_i= \frac{\gamma_{e}^{2}\hbar}
{4 \left(d^2+r_{i}^{2}\right)^{3/2}}
\left(1-3\cos^2{\psi_i}\right).
\label{deltai}
\end{equation}
$h_{\rm{dip}}$ is sensitive to the orientation of the
external magnetic field $\bm{B}=(\sin{\theta},0,\cos{\theta})B$ with
respect to the interface. This enters through 
\begin{equation}
\cos^2{\psi_i}=\frac{\left(d\cos{\theta}+r_i\cos{\phi_i}
\sin{\theta}\right)^2}{d^2+r_i^2}.
\end{equation}
For $\theta=0$, the average $h_{\rm{dip}}^{2}$ over an uniform DB area density
$\sigma_{{\rm DB}}$ is given by
\begin{eqnarray}
\langle h_{\rm{dip}}^{2}\rangle &=&\sigma_{{\rm DB}}\int_{0}^{2\pi} d\phi 
\int_{0}^{\infty}rdr 
h_{\rm{dip}}^{2}(r,\phi)\nonumber\\
&=& \frac{3\pi}{64}\sigma_{{\rm DB}} \frac{\gamma_{e}^{4} \hbar^2}{d^4}.
\label{avghdip}
\end{eqnarray}

\section{Hahn echo decay due to $1/f$ noise: Comparison to experiment\label{seconeoverf}}

The discussion above concluded that the following model for the
noise spectrum is valid at high temperatures ($k_BT\gg\delta$ and $k_BT\gg E_{\rm{max}}$):
\begin{equation}
\tilde{S}(\omega)=\left\{
\begin{array}{c c}
C/\omega_{\rm{min}}   & 0\leq |\omega|<\omega_{\rm{min}}\\
C/|\omega|  &
\omega_{\rm{min}} \leq |\omega| < \omega_{\rm{max}}\\
0 & \omega_{\rm{max}} \leq |\omega|<\infty
\end{array}
\right..
\label{oneof}
\end{equation}
The prefactor $C$ is given by 
\begin{equation}
  C=\frac{\langle h_{\rm{dip}}^{2}\rangle}{4\lambda_{\rm{max}}}\approx \frac{3\pi}{256}
\frac{\sigma_{\rm{DB}}}{\lambda_{\rm{max}}}\frac{\gamma_{e}^{4} 
    \hbar^2}{d^4}.
\label{Cexpression}
\end{equation}

We may calculate the Hahn echo response due to $1/f$ noise using
Eq.~(\ref{splus}) with the filter function Eq.~(\ref{fecho}).
If the inter-pulse time $\tau$ is neither too long (so that $c_{\rm{min}}=\pi
\tau \omega_{\rm{min}}/2<1$) nor too short (so that
$c_{\rm{max}}=\pi\tau
\omega_{\rm{max}}/2>1$)
we get 
\begin{eqnarray}
\left\langle\sigma_{+}(2\tau)\right\rangle&=&\exp{\bigg\{- C\tau^2 \bigg[
\bigg(4\ln{2}-\frac{2}{3}c_{\rm{min}}^{2}\bigg)}
-\frac{1}{4c_{\rm{max}}^{2}}\nonumber\\
&&\times\bigg(
3-4\cos{(2c_{\rm{max}})}+\cos{(4c_{\rm{max}})}
\bigg)
\bigg]\bigg\},
\label{vetauapprox}
\end{eqnarray}
after neglecting terms of order $c_{\rm{min}}^{3}$ and
$1/c_{\rm{max}}^{3}$. When $c_{\rm{min}} \lesssim 0.1$ and
$c_{\rm{max}}\gtrsim 10$ the echo envelope saturates and is well
approximated by the simpler expression
\begin{equation}
\left\langle\sigma_{+}(2\tau)\right\rangle\approx \exp{\left[
-4\ln{(2)}C\tau^2 
\right]},
\label{v_Esimple}
\end{equation}
that is independent of the low and high frequency plateaus assumed in
Eq.~(\ref{oneof}).\cite{note2,note1}

In the experiment of Ref.~\onlinecite{schenkel06}, each implanted Sb
donor is a probe of magnetic noise from the interface.  Because the
implanted profile is inhomogeneous, the parameter $C$ is different for
each layer of donors a distance $d$ below the interface.  The
experimental data was taken at $\delta/k_BT=0.3/5=0.06\ll 1$. From
Eq.~(\ref{Cexpression}) we obtain
\begin{equation}
\left\langle\sigma_{+}(2\tau)\right\rangle \approx 
\textrm{e}^{-\xi\left[\frac{2\tau}{\chi(d)}\right]^2},\label{vefit}
\end{equation}
\begin{equation}
\chi(d)=
\frac{6.25\;{\rm nm}}{\gamma_{e}^{2}\hbar}d^{2},
\end{equation}
\begin{equation}
\xi=\frac{\sigma_{\rm{DB}}\times (\rm{nm})^{2}}{\lambda_{\rm{max}}}.
\label{xi}
\end{equation}
In this approximation we may fit the experimental data using a single
dimensionless parameter $\xi$, provided the distribution of Sb donors
is well known.

\begin{figure}
\includegraphics[width=3in]{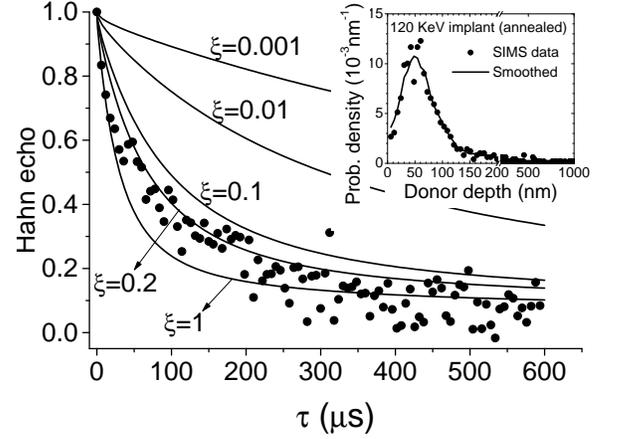}
\caption{Theoretical calculations (solid lines) and experimental data 
  (circles)\cite{schenkel06} for Hahn echo decay of Sb donors in the
  120~KeV implanted sample with the Si/SiO$_2$ surface. The theory is
  in reasonable agreement with the data when the theoretical parameter
  $\xi\approx 0.2$ [Eq.~(\ref{xi})]. The inset shows the Sb donor
  distribution measured by Secondary Ion Mass Spectroscopy (SIMS).}
\label{120KeVfit}
\end{figure}

\begin{figure}
\includegraphics[width=3in]{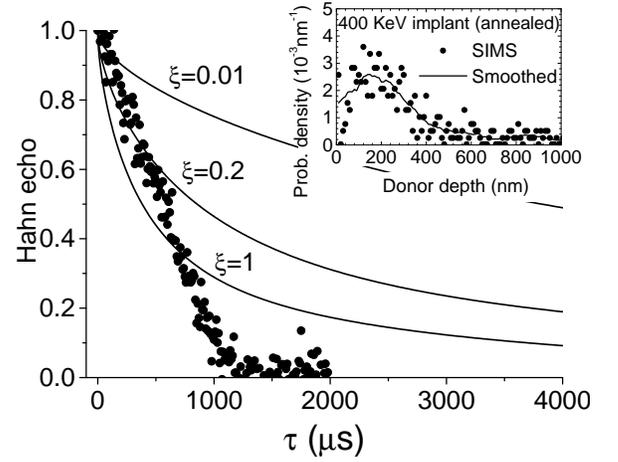}
\caption{Same as Fig.~\ref{120KeVfit} for the 400~KeV implanted
  sample, with the Si/SiO$_2$ surface. As pointed out in
  Ref.~\onlinecite{schenkel06}, the experimental data suffered from
  external field noise for $\tau>500$~$\mu$s.}
\label{400KeVfit}
\end{figure}

We used Eq.~(\ref{vefit}) together with the donor distribution
measured by Secondary Ion Mass Spectroscopy (SIMS) to obtain
theoretical estimates of Hahn echo decay relevant to the experiment of
Ref.~\onlinecite{schenkel06}.  Figs.~\ref{120KeVfit}~
and~\ref{400KeVfit} compares the theory with the 120~KeV and 400~KeV
implanted samples respectively, both with a Si/SiO$_2$ surface.  A
value of $\xi\approx 0.2$ for the theoretical curves seems to be
consistent with the experimental data. However, in the short time
range the theoretical curve seems to decay slower than the
experimental data, while at longer time intervals the theory seems to
decay faster. This lack of agreement may be due to deviations from the
measured SIMS distribution. The ultra-low donor densities were at the
sensitivity threshold for the SIMS technique, hence the donor
distribution is quite noisy [see insets of
Figs.~\ref{120KeVfit},~\ref{400KeVfit} - we used a numerically
smoothed version of the SIMS-annealed data of Figs.~1(a),~1(b) of
Ref.~\onlinecite{schenkel06}]. A higher probability density near the
interface could in principle explain the faster decay at shorter
times, while a deeper tail in the distribution could be responsible
for the slower decay at longer times.

The value for $\lambda_{\rm{max}}$ may be estimated from
$\lambda_{\rm{max}}=\frac{1}{2}\ln{\left(\frac{
      \bar{\Gamma}_{\rm{max}}}{\bar{\Gamma}_{\rm{min}}}\right)}\sim
\frac{1}{2}\ln{\left(\frac{10^6}{10^{-1}}\right)}\sim 10$. Combining
this with $\xi\sim 0.2$ we get $\sigma_{\rm{DB}}\sim 10^{14} {\rm
  cm}^{-2}$. 

We use Eq.~(\ref{v_Esimple}) and the value $\xi \approx 0.2$
extracted from experiment to estimate the
coherence time of a single donor located a distance $d$ below the
interface. This results in 
\begin{equation}
T_2(d)\approx 4\times 10^{-8}{\rm s} \left(\frac{d}{{\rm nm}}\right)^{2},
\label{t2estimate}
\end{equation}
with $T_2(d)$ inversely proportional to the square root of the DB area density.
The $1/f$ noise affecting a local magnetic probe a distance $d$ from
the interface is estimated as
\begin{equation}
\tilde{S}(\omega)\approx \left[6.5\times 10^{11}\rm{s}^{-2}
\left(\frac{10\;\rm{nm}}{d}\right)^{4}\right]\frac{1}{|\omega|},
\end{equation}
and is directly proportional to $\sigma_{\rm{DB}}$.

\section{Magnetic flux noise in SQUID devices\label{squid}}

The SQUID (superconducting quantum interference device) is probably
the most sensitive probe for magnetism at the nanoscale.  It consists
of a superconducting loop interrupted by two insulating barriers
(Josephson junctions). In this way it works as a magnetic
flux-to-voltage transducer.  SQUIDs are usually grown on top of a
Si/SiO$_2$ substrate, therefore magnetic noise due to dangling-bonds
within the SQUID loop will affect their performance as sensitive
magnetometers.

Our results on magnetic dipolar noise are easily translated to flux
noise in SQUIDs by substituting
$\langle h^{2}_{\rm{dip}}\rangle$ for
$\langle\Phi^{2}_{\rm{Total}}\rangle$ in
section~\ref{magnoise} above. In order to get an order of magnitude estimate for 
$\langle\Phi^{2}_{\rm{Total}}\rangle$, consider 
the flux produced by a single magnetic dipole moment $m_0$ located at
the center of a disk of radius $R$ (the area enclosed by the SQUID
loop).  In Gaussian units this is given by $\Phi_i=2\pi m_0 /R$. Each
dangling-bond contributes a dipole moment equal to
$m_0=\hbar\gamma_e/2=\hbar e /(2m_e c)$. Assuming an area density
$\sigma_{\rm{DB}}$ for the DBs leads to the following estimate for the
mean flux squared:
\begin{eqnarray}
\left\langle \Phi^{2}_{\rm{Total}}\right\rangle &\approx& \pi^3 \sigma_{\rm{DB}}\hbar^2 \gamma^{2}_{e}\nonumber\\
&=&2.49\times 10^{-11}\Phi_{0}^{2} \left[\sigma_{\rm{DB}}\times (\rm{nm})^{2}\right],
\label{flux2}
\end{eqnarray}
where $\Phi_0=hc/2e$ is the flux quantum.  The SQUID operates at very
low magnetic fields ($B\lesssim 1$~G), so the spin quantization
direction is set by local inhomogeneities and is different for each
DB.  The angular average of spin quantization direction reduces
Eq.~(\ref{flux2}) by a factor of $3$.  Moreover, taking account of
spins close to the superconducting wire and oriented along the SQUID
plane increases Eq.~(\ref{flux2}) by $\sim 3$.\cite{clarke07} As a
result, Eq.~(\ref{flux2}) has the same order of magnitude as the
calculation of Koch {\it et al} for loop sizes
$10-500$~$\mu$m.\cite{clarke07}

At high temperatures ($k_BT\gg \delta$ and $k_BT\gg E_{\rm{max}}$),
the flux noise due to the presence of DBs in the plane enclosed by
the SQUID loop is obtained by substituting $\langle
h^{2}_{\rm{dip}}\rangle \rightarrow \langle
\Phi^{2}_{\rm{Total}}\rangle$ in Eq.~(\ref{somegaegd}). The result is
\begin{equation}
\tilde{S}_{\Phi}(\omega)=
\left[\frac{\sigma_{\rm{DB}}\times (\rm{nm})^{2}}{\lambda_{\rm{max}}}\right]
6.2\times 10^{-12}
\frac{\Phi^{2}_{0}}{|\omega|}.
\label{noisesquid}
\end{equation}
The value in brackets equals the parameter $\xi$ used to fit our ESR
experiment (Fig.~\ref{120KeVfit}).  Using $\xi \approx 0.2$ we get an
estimate for the flux noise contribution from an untreated Si/SiO$_2$
substrate:
\begin{equation}
\tilde{S}_{\Phi}(\omega)\approx 1.2\times 10^{-12} \frac{\Phi^{2}_{0}}{|\omega|}.
\label{squidestimate}
\end{equation}
Interestingly, this result has the same order of magnitude as the measurements of 
Ref.~\onlinecite{yoshihara06} using a small flux qubit as
a probe of magnetic noise. A compilation of measurements of flux noise
in SQUIDs was given recently by Ref.~\onlinecite{clarke07}, where we
see that $\tilde{S}(1\textrm{Hz})$ lies in the range $0.1-100\times
10^{-12}\Phi_{0}^{2}$ for a wide variety of samples. 

Note that the high temperature condition $k_BT\gg \delta$ implies
$T\gg 0.1$~mK for the low magnetic fields ($\sim 1$~G) in SQUIDs.
Unfortunately, there are no estimates of $E_{\rm{max}}$ for a
Si/SiO$_2$ interface. For bulk SiO$_2$ the values $E_{\rm{max}}\sim
10$~K and $E_{\rm{min}}\sim 0.1$~K were estimated.\cite{galperin89} We
emphasize that Eq.~(\ref{squidestimate}) is the maximum value for the
noise, which saturates at $k_BT\gg E_{\rm{max}}$. For
$k_BT<E_{\rm{max}}$, Eq.~(\ref{squidestimate}) will be reduced by a
factor $(K_BT/E_{\rm{max}})^{1+\alpha}$, and the frequency dependence
will change to $1/|\omega|^{p}$ with $p=1.2-1.5$, see
Eq.~(\ref{somegap}).

\section{Nuclear spin noise from a hydrogen passivated surface\label{nuclearsd}}

Surface passivation with hydrofluoric acid drastically reduces the
amount of dangling-bonds. Nevertheless this occurs at
the expense of adding a large amount of hydrogen nuclear spins.  Here
we investigate the magnetic noise mechanism arising due to the dipolar
fluctuation of hydrogen nuclear spins at a perfect passivated Si-H
surface.

It is well established that spin decoherence of donors in bulk natural
silicon is dominated by nuclear spin noise from the 4.67\% $^{29}$Si
nuclear spins.\cite{witzel05,desousa06} The samples studied
here\cite{schenkel06} are known to have less than 0.1\% of $^{29}$Si
isotopes, leading to a contribution of the order of $\frac{1}{T_2}\sim
10$~Hz. For a Si/SiO$_2$ sample, the fraction of oxygen isotopes with
non-zero nuclear spin is even lower (0.038\%), hence oxygen nuclear
spins should be a minor contributor to magnetic noise at oxidized
samples.

\begin{figure}
\includegraphics[width=3.4in]{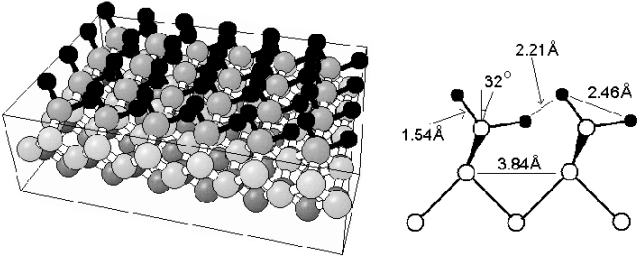}
\caption{A Hydrogen terminated silicon surface is obtained after
  immersing an oxidized sample in a hydrofluoric acid solution. Here
  we show a Si(100)H surface with the hydrogen atoms forming a
  canted-row dihydride structure.\cite{northrup91} The SiH$_{2}$
  groups form a square lattice of side $5.43/\sqrt{2}=3.84$~\AA.}
\label{fig_Si100H}
\end{figure}

We carry out a model calculation for the Si(100)H surface under the
assumption that the hydrogen atoms are arranged in a canted-row
dihydride phase with no orientation disorder, see
Fig.~\ref{fig_Si100H}.\cite{northrup91} The truncated Hamiltonian for
a single donor electron spin interacting with the hydrogen nuclear
spin lattice at the surface is given by
\begin{eqnarray}
{\cal H}&=&\frac{1}{2}\gamma_e B \sigma_z -\gamma_n B \sum_i I_{iz} + 
\frac{1}{2}\sum_i D_{i} I_{iz}\sigma_z\nonumber\\
&&+\sum_{i\neq j} b_{ij} \left(
I_{i+}I_{j-} -2I_{iz}I_{jz}\right),
\label{hnuc}
\end{eqnarray}
where $\bm{I}_{i}$ is the nuclear spin operator for the hydrogen atom
located at position $\bm{R}_i$ with respect to the electron
[$\gamma_n=2.66\times 10^{4}$~(sG)$^{-1}$ is the gyromagnetic ratio
for the hydrogen nuclear spin]. Note that in Eq.~(\ref{hnuc}) we have
neglected the non-secular contribution of the electron-nuclear dipolar
interaction. This approximation is valid only at higher external
magnetic fields, $\gamma_e B\gg \sqrt{\sum_{i} D_{i}^{2}}$. For
$d=10$~nm, $B>0.1$~T is necessary to satisfy this criteria.  One can
show numerically that the non-secular interactions produce a loss of
visibility for the Hahn echo envelope scaling as $\sim\sum_{i}
D_{i}^{2}/(\gamma_e B)^2$.\cite{shenvi05} The electron-nuclear dipolar
coupling is given by
\begin{equation}
D_{i}=\frac{\gamma_n\gamma_{e}\hbar}{R_{i}^3}\left(1-3\cos^2\theta_i \right),
\end{equation}
where $\theta_i$ is the angle between ${\bf R}_i$ and the direction of
the external magnetic field. Each pair of hydrogen nuclear spins
labeled by $i,j$ are mutually coupled by the dipolar
interaction
\begin{equation}
b_{ij}=-\frac{1}{4}\frac{\gamma^{2}_{n}\hbar}{R_{ij}^3}
\left(1-3\cos^2\theta_{ij}\right),
\end{equation}
where $\theta_{ij}$ is the angle between the $B$ field and the vector
$\bm{R}_{ij}$ linking the two nuclear spins. 

The Hamiltonian Eq.~(\ref{hnuc}) is directly mapped into the effective
model Eq.~(\ref{hmodel}) through the prescription $\eta_z=\sum_{i}D_{i}
I_{iz}$. The noise spectrum [Eq.~(\ref{noisedef})] is then calculated
using a ``flip-flop'' approximation, i.e. assuming a model that
considers only flip-flop transitions between pairs of nuclear spins.
In this approximation, the noise spectrum becomes\cite{desousa06}
\begin{equation}
\tilde{S}(\omega)=\sum_{i<j}\frac{b_{ij}^2 \Delta_{ij}^2}
{b_{ij}^2+\Delta_{ij}^2}
\left[\delta(\omega+E_{ij})+\delta(\omega-E_{ij})\right],
\label{noisedelta}
\end{equation}
with $\Delta_{ij}=(D_{i}-D_{j})/4$, and $E_{ij}=2\sqrt{b_{ij}^{2}
  +\Delta_{ij}^{2}}$. We dropped the inhomogeneous broadening term
proportional to $\delta(\omega)$ in Eq.~(\ref{noisedelta}) because it
does not contribute to Hahn echo decay. 

Using Eq.~(\ref{splus}) and
Eq.~(\ref{noisedelta}) the Hahn echo envelope becomes
\begin{equation}
\left\langle\sigma_{+}(2\tau)\right\rangle
=\prod_{i<j} \textrm{e}^{-4b_{ij}^{2}\Delta_{ij}^{2}\tau^4 {\rm sinc}^4\left(
\tau\sqrt{b_{ij}^2+\Delta_{ij}^2}
\right)},
\label{vqt}
\end{equation}
where ${\rm sinc}\,x =\sin{x}/x$. This result is identical to the
lowest order cluster expansion derived in Ref.~\onlinecite{witzel05}
through direct calculation of the spin echo response. Another way to
derive Eq.~(\ref{vqt}) is to assume that the nuclear spin pair
transitions are quasiparticle excitations with infinite
lifetime.\cite{yao06} Eq.~(\ref{vqt}) is able to predict the Hahn echo
decay without any phenomenological fitting parameter, in contrast to
the traditional ``Brownian motion'' models developed
previously.\cite{klauder62}

Note that the magnetic noise due to nuclear spins is a linear
combination of sharp peaks (delta functions), reflecting the
mesoscopic nature of the nuclear spin bath. Each delta function is a
transition between discrete nuclear spin energy levels. This is in
contrast to the continuous (Lorentzian) noise due to a single
dangling-bond spin interacting with the phonon continuum. 

In order to plot a continuous noise spectrum we represent the delta
functions in Eq.~(\ref{noisedelta}) by normalized Gaussians with
linewidth $\sigma=10^2$~s$^{-1}$. Note that the Hahn echo decay is
independent of the particular choice of $\sigma$ or the Gaussian
lineshape provided $\tau$ remains much smaller than $1/\sigma$ [in
this case the Hahn echo envelope calculated by
Eqs.~(\ref{splus}),~(\ref{fecho}) with a coarse grained noise spectrum
is very well approximated by the zero broadening expression
Eq.~(\ref{vqt}).  

Fig.~\ref{noise_sih} shows the nuclear spin noise spectrum from the
point of view of a probe (a donor spin) lying $10$~nm below the
surface.  Interestingly, we find that the noise spectrum is
sensitive to the relative orientation of the external magnetic field
with respect to the surface. The noise has a global minimum for
$\theta\approx 50^{\circ}$. As shown in Fig.~\ref{t2_angle}, this
effect translates into a variation of about 50\% in the electron spin
decoherence time $T_2$ [$T_2$ is obtained as the $1/e$ decay of the
Hahn echo given by Eq.~(\ref{vqt})].  This orientation dependence is
surprisingly different than the one in bulk Si:P, see e.g. Fig.~8 of
Ref.~\onlinecite{desousa06}.  Fig.~\ref{t2_angle} shows that $T_2$ is
minimized when $\theta=0$ and maximized when $\theta\approx
50^{\circ}$, in contrast to bulk Si:P where precisely the opposite was
found. This special orientation dependence is the fingerprint of
nuclear spin noise in a Si(100)H surface, allowing a clear
identification of this mechanism in pulse spin resonance.

\begin{figure}
\includegraphics[width=3in]{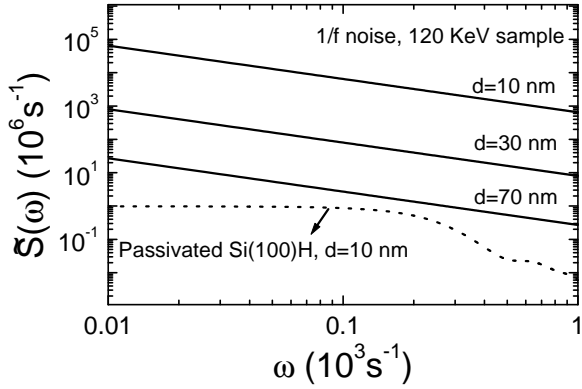}
\caption{Magnetic noise spectrum for the oxidized interface as probed by a single donor spin 
  a distance $d$ below the interface. For comparison, we show the
  nuclear spin noise spectrum in a hydrogen passivated surface.}
\label{comp_nois}
\end{figure}

\begin{figure}
\includegraphics[width=3in]{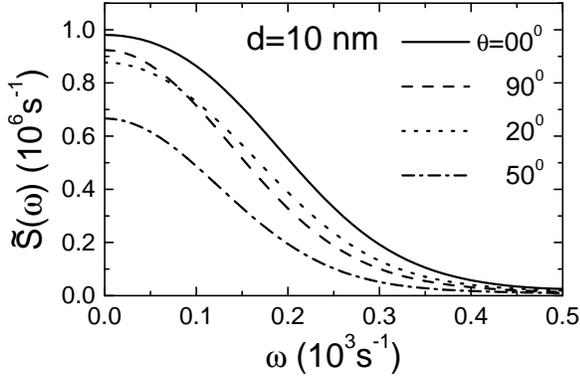}
\caption{Magnetic noise spectrum due to the Si(100)H surface as probed
  by a single donor spin 10~nm below the surface. We show the noise
  spectrum for four different angles $\theta$, labeling the relative
  orientation of the external magnetic field with respect to the
  surface.}
\label{noise_sih}
\end{figure}

\begin{figure}
\includegraphics[width=3in]{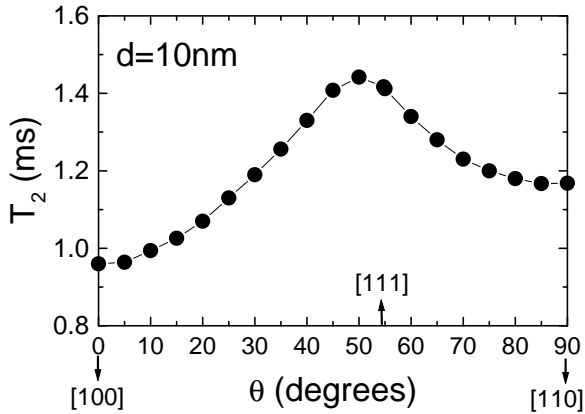}
\caption{Magnetic field angular dependence of $T_2$ for a donor spin located
  10~nm below a hydrogen terminated surface. $\theta$ is the angle
  between the external magnetic field and the (100) direction. The
  resulting orientation dependence is quite distinct from the one due
  to $^{29}$Si nuclear spins in bulk natural
  silicon.\cite{desousa06,tyryshkin06}}
\label{t2_angle}
\end{figure}

\begin{figure}
\includegraphics[width=3in]{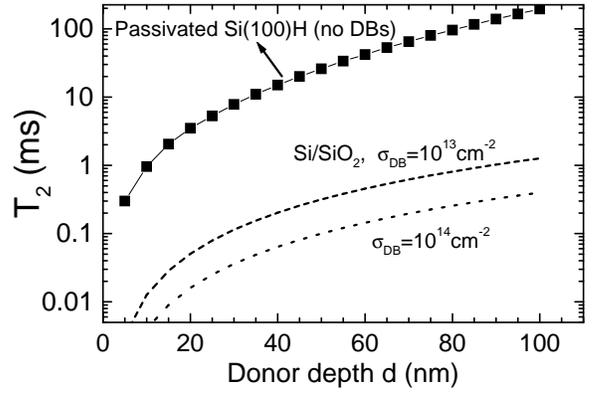}
\caption{Spin decoherence time $T_2$ as a function of the donor 
  distance from the surface, for a passivated Si(100)H surface
  containing no dangling-bonds (squares) and for a Si/SiO$_2$
  interface containing a dangling-bond density equal to
  $10^{14}$~cm$^{-2}$ [Eq.~(\ref{t2estimate})].}
\label{t2_depth}
\end{figure}

Fig.~\ref{t2_depth} shows $T_2$ as a function of the donor distance
from the surface. Note that we find $T_2>10$~ms for $d\sim 30$~nm,
suggesting that this mechanism should not be playing a dominant role
in the shallow implanted sample of Ref.~\onlinecite{schenkel06}
(120~KeV sample). For $d>100$~nm, $T_2$ is hundreds of milliseconds,
so hydrogen nuclear spins are not affecting the 400~KeV implanted
sample either. The nuclear spin noise in a passivated surface may be
further reduced by a factor of $\sim 4$ by using deuterium instead of
hydrogen (the deuterium gyromagnetic ratio is $3.28$ times smaller
than hydrogen). This results in donor $T_2$'s greater by a factor of
two.

For a perfect hydrogen passivated surface the theoretical $T_2$'s are
much longer than the values measured in Ref.~\onlinecite{schenkel06}.
It is well known that chemical passivation of a Si(100) surface can
not remove all dangling-bonds, in contrast to Si(111) that usually
removes nearly all dangling-bonds.\cite{yablonovitch86} Therefore the
dangling-bond mechanism might still be playing a role in the
passivated samples. Repeating the experiment for the Si(111) surface
could possibly yield even longer coherence times.  The finite Sb
density in these samples implies that the mutual interaction between
donor spins (donor-donor dipolar coupling) might play a role, a
mechanism of decoherence referred to as ``instantaneous
diffusion''.\cite{raitsimring74} We have confirmed this expectation by
showing that the contribution to $T_2$ due to instantaneous diffusion
is of the order of 0.3 and 1 millisecond for the 120~KeV and 400~KeV
samples respectively.  Therefore instantaneous diffusion might explain
a fraction of the measured echo decay rates.
Refs.~\onlinecite{raitsimring74,tyryshkin03} discusses a method for
completely removing the instantaneous diffusion mechanism in a doped
sample.

With respect to SQUID devices, we remark that the nuclear spin
flip-flop mechanism considered in this section does not contribute to
magnetic flux noise (a flip-flop preserves the value of the magnetic
moment for two nuclear spins, leaving the total flux unchanged).  The
statistical fluctuation of individual hydrogen nuclear spins (due to a
finite $T_{1}^{H}$) should be extremely small because $T_{1}^{H}$ is
usually hundreds of seconds or more. The nuclear spin noise due to
ensemble fluctuations of nuclear spins may be detected by SQUIDs under
optimal circumstances, see Ref.~\onlinecite{sleator85}.

\section{Discussion}

In summary, we developed a theory of magnetic noise due to spin-flips
of paramagnetic centers at an amorphous semiconductor-oxide interface.
The mechanism of dangling-bond spin relaxation due to its interaction
with tunneling-two-level systems and phonons of the amorphous
interface was discussed in detail. We also showed how these effects
may be greatly reduced by surface passivation with hydrogen.
Substituting the paramagnetic dangling-bonds with a monolayer of
hydrogen nuclear spins reduces the magnetic noise level by many orders
of magnitude, as seen in Fig.~\ref{comp_nois}.  We related these
results to decoherence of spin qubits in silicon as a function of
their distance from the interface and flux noise in SQUID qubits.

Our work generalizes and extends the model of dangling-bond
spin-lattice relaxation in amorphous materials originally proposed in
Refs.~\onlinecite{kurtz80,lyo80,askew86}.  Particularly, we clarified
the different temperature and magnetic field dependence as a function
of the ratio between TTLS energy splitting $E$ and DB spin Zeeman
energy $\delta$.  

The theory of paramagnetic DB spin relaxation is significant
for two recent proposals of single spin measurement based on
spin-dependent recombination of conduction electrons with
dangling-bonds close to the Fermi level.\cite{xiao04,boehme06} In
these experiments the time scale $T_{1}^{{\rm DB}}$ sets the limit on
single spin measurement fidelity. To our knowledge there is yet no
experimental study of $T_{1}^{{\rm DB}}$ at the Si/SiO$_2$ surface. We
propose the measurement of magnetic field and temperature dependence
of DB spin relaxation at short times [Eq.~(\ref{t1dblow})] and the
non-exponential decay at longer times [Eq.~(\ref{nonexponential})] in
order to validate our theoretical results and give a full
characterization of the free parameters.

Our calculations provide benchmark values for the ultimate coherence
times of group V donor spin qubits implanted in an actual device
structure made from nuclear-spin-free silicon.  Although the longest
coherence times are in principle achievable with a perfect oxidized
surface \emph{without dangling-bonds}, the inevitable presence of a
large density of these defects in real devices make surface
passivation an attractive alternative. Since each donor must be
positioned close to an insulating interface in order to allow gate
control of exchange,\cite{kane98} hyperfine
couplings,\cite{kettle03,martins04} as well as electron
shuttling,\cite{skinner03} the interface effects described here will
play an important role in the material optimization of silicon devices
exploiting spin coherence.

Recently, $^{29}$Si nuclear magnetic resonance experiments in
polycrystalline silicon at room temperature were interpreted using a
model of magnetic $1/f$ noise.\cite{ladd05} The proposed mechanism was
related to the charge fluctuation of trapping-centers at the surface
of the microcrystals.  Our work suggests that it is the spin-flip of
paramagnetic DBs, not trapping-centers, that probably account for most
of the $1/f$ noise observed in Ref.~\onlinecite{ladd05}.

Koch {\it et al.} proposed a model of $1/f$ flux noise in SQUIDs based
on electron hopping to localized defect sites, and concluded that a
quite high trapping-center area density ($5\times 10^{13}$~cm$^{-2}$)
was required to explain flux noise in SQUID qubits.\cite{clarke07} Our
work suggests that the spin-flip of paramagnetic centers from the
substrate may provide an alternative explanation, based on a more
physical \emph{paramagnetic dangling-bond density} similar to the one
estimated in their work.

We remark that a $C-V$ analysis of an unannealed Si/SiO$_2$ interface
leads to an energy density equal to $\rho'\sim
10^{13}$~eV$^{-1}$cm$^{-2}$ (See Fig.~4 of
Ref.~\onlinecite{gerardi86}).
This implies that the SQUID substrate is contributing at most
$k_BT\rho'\sim 10^{10}$~cm$^{-2}$ of trapping-center area density at
$T=0.1-4$~K.  Nevertheless, the area density for paramagnetic DBs
should correlate with $U\rho'\sim 10^{13}$~cm$^{-2}$. The value
obtained here ($10^{14}$~cm$^{-2}$) is a factor of $10$ higher.

STM experiments provide another way to estimate the trap energy
density.  In Ref.~\onlinecite{koch87}, a clean Si(100) surface was
exposed to low pressure oxygen in order to produce approximately a
single oxygen monolayer. When the tip to surface voltage was $\sim
1$~V, ten to one hundred trapping-centers could be detected in a
$65\times 65$~\AA$^{2}$ region. This leads to an energy density in the
range $10^{13}-10^{14} ~\rm{eV}^{-1}\rm{cm}^{-2}$, higher than the
$C-V$ measurements.

The frequency and temperature dependence of flux noise in SQUIDs was
measured a while ago in Ref.~\onlinecite{wellstood87}, using a wide
variety of samples.  These included silicon oxide substrates deliberately
and not deliberately oxidized, as well as sapphire substrates. Some
samples showed no temperature dependence, and the frequency dependence
appeared to fit 1/$f^{p}$ , with $p=0.58-0.80$.  This frequency
dependence can not be explained by our model.  Nevertheless, the
absence of temperature dependence may be explained by our model,
provided the majority of DBs are connected to one or more thermally
activated TTLS (or equivalently, $E_{\rm{max}}<k_BT$).  An interesting
question for future research is whether the interaction between DB
spins can account for this discrepancy.

This work establishes an important connection between flux noise in
SQUID devices and ESR studies of implanted donor impurities or
dangling-bonds.  As a result, ESR characterization may play an
important role in the pre-screening of novel materials for SQUID
fabrication.

\acknowledgements

We acknowledge useful discussions with C. Boehme, J. Clarke, P.M.
Lenahan, and T.C. Shen. We are particularly grateful to J. Bokor, C.C.
Lo, S.A.  Lyon, T. Schenkel, and A.M. Tyryshkin for discussions and
for providing the experimental data presented in this paper.  We
acknowledge financial support from the Nanoelectronics Research
Initiative (NRI) - Western Institute of Nanoelectronics (WIN), the
National Security Agency (NSA) under MOD 713106A, NSF under Grant No.
0404208 and by the Department of Energy under Contract No.
DE-AC03-76SF00098.

\appendix

\section{Magnetic 1/$f$ noise due to dangling-bond charge
  trapping-centers\label{trappingcenters}}

Here we consider a mechanism of magnetic noise similar to the
McWhorter model of current noise in semiconductor devices, that does
not involve phonons.\cite{kogan96}

Dangling-bonds with energy close to the Fermi level act as charge
trapping-centers, capturing electrons from interface states.  
Magnetic noise occurs because each time an
interface electron tunnels into the DB, it produces an effective
dipolar field in the donor spin, that is given by Eq.~(\ref{deltai}) divided
by two. 

The tunneling rate for a trap located a distance $z$ from the
interface is assumed to be $\Gamma(z)=\Gamma_{{\rm
    max}}\textrm{e}^{-z/\lambda}$.  Here $z$ measures the depth of the
charge trap into the SiO$_2$ dielectric, $z=0$ refers to a trap at the
Si/SiO$_2$ interface. We have
$\lambda=\sqrt{\hbar^2/(2m^{*}E_0)}/2\sim 1$~\AA, where $E_0\approx
4$~eV is half the gap difference between the two materials.
$\Gamma_{{\rm max}}\sim 10^9$~s$^{-1}$ depends on the cross section
for electron capturing.  Assuming a uniform distribution for $z$ in
the interval $[0,z_{{\rm max}}]$ leads to a distribution of rates
equal to
\begin{equation}
p(\Gamma)=\frac{p(z)}{|\frac{d\Gamma}{dz}|}\approx
\frac{\lambda/z_{{\rm max}}}{\Gamma},
\end{equation}
for $\Gamma_{{\rm min}}<\Gamma<\Gamma_{{\rm max}}$, and zero
otherwise. The noise spectrum is readily calculated as
\begin{eqnarray}
\tilde{S}(\omega)&=&\frac{1}{4}\left\langle h_{{\rm
      dip}}^{2}\right\rangle
\int_{\Gamma_{{\rm min}}}^{\Gamma_{\rm max}}d\Gamma p(\Gamma)
      \frac{\Gamma/\pi}{\omega^2+\Gamma^2}\nonumber\\
&=&\frac{3\pi}{512}
\frac{\lambda}{z_{{\rm max}}}
(\rho'k_BT) \frac{\gamma_{e}^{4} \hbar^2}{d^4}\frac{1}{\omega}.
\end{eqnarray}
Therefore we have $C/\omega$ noise with parameter $C=\lambda\langle
h_{{\rm dip}}^{2}\rangle/(8z_{{\rm max}})$. In order to compare to
Hahn echo decay data we define a parameter $\xi$ similar to
Eq.~(\ref{xi}). This is
given by 
\begin{equation}
\xi=\frac{\lambda}{2z_{{\rm max}}}
\rho'k_BT({\rm nm}^2).
\end{equation}
The area density for DB's with energy close to the Fermi level is
estimated as $\rho'k_BT$, with $\rho'\sim
10^{13}$~cm$^{-2}$eV$^{-1}$.\cite{gerardi86} The maximum possible
value of $\xi$ (at $T=5$~K) is estimated from $\rho'k_BT<
10^{10}$~cm$^{-2}$.  Assuming $z_{{\rm max}}\sim 10\lambda$ we get
$\xi< 10^{-5}$. We remark that this maximum possible value for $\xi$
is four orders of magnitude smaller than the value required to explain
the experimental data of Ref~\onlinecite{schenkel06} (see
Figs.~\ref{120KeVfit}~and~\ref{400KeVfit}).

\section{Calculation of the DB spin time-dependent correlation function\label{calctimecor}}

We are concerned with finite frequency fluctuations of the DB spin
magnetic moment along the $B$ field (z direction), therefore we will describe
the dissipative kinetics of the DB+TTLS network considering only diagonal
density matrix elements in the Bloch-Wangsness-Redfield theory.

Define the propagator matrix $P_{ij}(t)$ as the diagonal density
matrix element $\rho_{ii}(t)$ subject to the initial condition
$\rho_{lm}(0)=\delta_{lj}\delta_{mj}$. This is just the probability
that the DB+TTLS will be at the state $i$ at time $t$ given that it was at
state $j$ at time $t=0$. Note that here the indexes $i$ labels one of the
four DB+TTLS levels $(+\downarrow,-\downarrow,+\uparrow,-\uparrow)$ [See Fig.~\ref{fig3}(b)]. 
Furthermore,
define the matrix $p_{ij}(t)=P_{ij}(t)-w_{i}$, with
$w_{i}=\rho_{ii}(\infty)$ the equilibrium probabilities for level $i$.
The steady state solution is then given by
$p_{ij}(t\rightarrow\infty)=0$.  With this definition the
rate equations for the $4\times 4$ matrix $\bm{p}$
becomes simply
\begin{equation}
\frac{d}{dt}\bm{p}(t)=-\bm{\Lambda}\cdot \bm{p}(t),
\label{ddt}
\end{equation}
with initial condition $p_{ij}(0)=\delta_{ij}-w_i$, and a relaxation
tensor $\bm{\Lambda}$ defined as follows: For $i\neq k$,
$\Lambda_{ik}=-\Gamma_{k\rightarrow i}$, that's
minus the rate for entering level $i$
from level $k$.  For $i=k$, $\Lambda_{ii}=\sum_{j(\neq i)}\Gamma_{i\rightarrow j}$,
or the sum of rates for exiting
level $i$. The case of
our four-level system without direct transitions between Zeeman sub-levels
is described by
\begin{equation}
\bm{\Lambda}=\left(
\begin{array}{cccc}
r_{+} + \Gamma_{+\downarrow}& -r_{-}& 0&
 -\Gamma_{-\uparrow}\\
-r_{+} & r_{-}+\Gamma_{-\downarrow}&
-\Gamma_{+\uparrow}& 0\\
0 & -\Gamma_{-\downarrow}& r_{+}+
\Gamma_{+\uparrow}& -r_{-}\\
-\Gamma_{+\downarrow} & 0& -r_{+}&
r_{-}+\Gamma_{-\uparrow} 
\end{array}
\right).\label{lambmat}
\end{equation}
The vector for equilibrium probabilities
$\bm{w}=(w_{+\downarrow},w_{-\downarrow},w_{+\uparrow},w_{-\uparrow})$
is the eigenvector of $\Lambda$ with eigenvalue zero, satisfying
$\sum_i w_i=1$.

In the four-level system notation
$(+\downarrow,-\downarrow,+\uparrow,-\uparrow)$ the vector $\bm{x}$ of
dipolar fields assumes the values $\bm{x}=h_{\rm{dip}}(-1,-1,+1,+1)$.
The correlation function becomes
\begin{eqnarray}
S_z(t)&=&\left\langle \left[\eta_z(t)-\bar{\eta}_z\right]
\left[\eta_z(0)-\bar{\eta}_z\right]\right\rangle\\
&=&\sum_{i,j}x_i\left[P_{ij}(t)-w_i\right]x_j w_j\\
&=&\bm{x}\cdot \bm{p}(t)\cdot \bm{x_{w}},\label{stexp1}
\end{eqnarray}
with $\bm{x_{w}}=(x_1 w_1,x_2 w_2,\ldots)$. Eq.~(\ref{stexp1}) together
with its explicit solution $\bm{p}(t)=\textrm{e}^{-\bm{\Lambda}t}$
allows exact calculations of the correlation function $S_z(t)$.  

%
%


\begin{thebibliography}{99}
\bibitem{note3} The local electric field $\bm{E}$ has contributions
  from the charge accumulation layer at the interface, as well as from
  ligands and atomic distortions. Hence in general $\bm{E}$ has a
  component along the interface, which adds a term proportional to
  $S_{z}^{\rm{DB}}$ to Eq.~(\ref{ttlsspindiag}). This is found to
  produce a small shift in the DB $g$-factor, with no qualitative
  effect on $T^{\rm{DB}}_{1}$.
\bibitem{note0} Note that the short time behavior must be extracted
  directly from $\left\langle S^{\rm{DB}}_{z}(t)\right\rangle =\sum_i
  \textrm{e}^{-\Gamma_i |t|}$, which is a non-analytic function at
  $t=0$. Expanding at $t=0+$ gives the correct linear decay. A
  different result (quadratic) is obtained from expanding the inverse
  Fourier transform of Eq.~(\ref{somegaegd}): This is incorrect
  because $\left\langle S^{\rm{DB}}_{z}(t)\right\rangle$ is not
  analytic at $t=0$.
\bibitem{note2} The analogous expression for free induction decay
  (FID) due to $1/f$ noise is equal to 
\begin{equation}
\left\langle\sigma_{+}(t)\right\rangle\approx 
\exp{\left\{-Ct^2\left[
0.9328-\ln{\left(\omega_{\rm{min}}t\right)}
\right]\right\}}.\nonumber
\end{equation}
The presence of a logarithmic singularity in the exponent implies that
FID is usually sensitive to the particular value chosen for the low
frequency cut-off $\omega_{\rm{min}}$. This is not the case for Hahn
echo decay.
\bibitem{note1} If $c_{\rm{min}}$ or $c_{\rm{max}}$ are outside the range of validitity
  for Eq.~(\ref{v_Esimple}), we may still use Eq.~(\ref{v_Esimple}) as
  a lower bound on the coherence (i.e. the $\approx$ can be
  substituted by a $>$).
\end{thebibliography}
\end{document}